\documentclass {elsart}
\usepackage{algorithm}
\usepackage{algorithmic}
\usepackage{graphicx}

\usepackage{graphics}
\usepackage{graphicx}
\usepackage{epsfig}

\usepackage{amssymb}

\begin{document}

\begin{frontmatter}



\title{Generalized Statistics Framework for Rate Distortion Theory}


\author[SRC]{R. C. Venkatesan\corauthref{cor}}
\corauth[cor]{Corresponding author.}
\ead{ravi@systemsresearchcorp.com}
\author[UNLP]{A. Plastino}
\ead{plastino@venus.fisica.unlp.edu.ar}

\address[SRC]{Systems Research Corporation,
Aundh, Pune 411007, India}
\address[UNLP]{IFLP, National University La Plata \&
National Research Council (CONICET)\\ C. C., 727 1900, La Plata,
Argentina}

\begin{abstract}
Variational principles for the rate distortion (RD) theory in lossy
compression are formulated within the ambit of the generalized
nonextensive statistics of Tsallis, for values of the nonextensivity
parameter satisfying $ 0 < q < 1 $ and $ q
> 1 $.   Alternating minimization numerical schemes
to evaluate the nonextensive RD function, are derived. Numerical
simulations demonstrate the efficacy of generalized statistics RD
models.
\end{abstract}

\end{frontmatter}

\section{Introduction}
The generalized (nonadditive) statistics of Tsallis' [1,2] has
recently been the focus of much attention in statistical physics,
and allied disciplines.  Nonadditive statistics\footnote{The terms
generalized statistics, nonadditive statistics, and nonextensive
statistics are used interchangeably.}, which generalizes the
extensive Boltzmann-Gibbs-Shannon statistics, has much utility in a
wide spectrum of disciplines ranging from complex systems and
condensed matter physics to financial mathematics \footnote{A
continually updated bibliography of works related to nonextensive
statistics may be found at http://tsallis.cat.cbpf.br/biblio.htm.}.
This paper investigates nonadditive statistics within the context of
Rate Distortion (RD) theory in lossy data compression.

RD theory constitutes one of the cornerstones of contemporary
information theory [3, 4], and is a prominent example of
\textit{source coding}. It addresses the problem of determining the
minimal amount of entropy (or information) $R$ that should be
communicated over a channel, so that a compressed (reduced)
representation of the source (input signal) can be approximately
reconstructed at the receiver (output signal) without exceeding a
given distortion $D$.

  For a thorough exposition of RD
theory Section 13 of  [3] should be consulted. Consider a discrete
random variable $ X \in \mathcal{X} $\footnote{Calligraphic fonts
denote sets.}called the \textit{source alphabet} or the
\textit{codebook}, and, another discrete random variable $ \tilde
X \in \tilde \mathcal{X} $ which is a compressed representation of
$ X $.  The compressed representation $ \tilde X $ is sometimes
referred to as the \textit{reproduction alphabet} or the
\textit{quantized codebook}. By definition, quantization is the
process of  approximating a continuous range of values (or a very
large set of possible discrete values) by a relatively small set
of discrete  symbols or integer values.

The mapping of $ x \in  \mathcal{X} $ to $ \tilde x \in \tilde
\mathcal{X} $ is characterized by a conditional (transition)
probability $ p(\tilde x|x) $.  The information rate distortion
function is obtained by minimizing the generalized mutual entropy
$ I_{q}(X;\tilde X) $ (defined in Section 2)~\footnote{The absence
of a principled nonextensive channel coding theorem prompts the
use of the term mutual entropy instead of mutual information.}
over all normalized $ p\left( {\tilde x\left| x \right.} \right)
$. Note that in RD theory $ I_q(X;\tilde X) $ is known as the
\textit{compression information} (see Section 4). Here, $ q $ is
the nonextensivity parameter [1, 2] defined in Section 2.

RD theory has found applications in diverse disciplines, which
include data compression and machine learning. Deterministic
annealing [5,6] and the information bottleneck method [7] are two
influential paradigms in machine learning, that are closely related
to RD theory. The representation of RD theory in the form of a
variational principle, expressed within the framework of the Shannon
information theory, has been established [3]. The computational
implementation of the RD problem is achieved by application of the
Blahut-Arimoto alternating minimization algorithm [3,8], derived
from the celebrated Csisz\'{a}r-Tusn\'{a}dy theory [9].

Since the work on nonextensive source coding by Landsberg and Vedral
[10], a number of studies on the information theoretic aspects of
generalized statistics pertinent to coding related problems have
been performed by Yamano [11], Furuichi [12, 13], and Suyari [14],
amongst others. The source coding theorem, central to the RD
problem, has been derived by Yamano [15] using generalized
statistics. A preliminary work by Venkatesan [16] has investigated
into the re-formulation of RD theory and the information bottleneck
method, within the framework of nonextensive statistics.

Generalized statistics has utilized a number of constraints to
define expectation values.  The linear constraints originally
employed by Tsallis of the form $ \left\langle A \right\rangle =
\sum\limits_i {p_i } A_i $ [1], were convenient owing to their
similarity to the maximum entropy constraints.  The linear
constraints were abandoned because of difficulties encountered in
obtaining an acceptable form for the partition function.  These were
subsequently replaced by the Curado-Tsallis (C-T) [17] constraints $
\left\langle A \right\rangle _q  = \sum\limits_i {p_i^q } A_i  \ $.
The C-T constraints were later discarded on physics related grounds,
$ \left\langle 1 \right\rangle _q \ne 1 $, and replaced by the
normalized Tsallis-Mendes-Plastino (T-M-P) constraints [18] $
\left\langle {\left\langle A \right\rangle } \right\rangle _q  =
\sum\limits_i {\frac{{p_i^q }}{{\sum\limits_i {p_i^q } }}A_i } \ $.
The dependence of the expectation value on the normalized
\textit{pdf} renders the canonical probability distributions
obtained using the T-M-P constraints to be
\textit{self-referential}.  A fourth form of constraint, prominent
in nonextensive statistics, is the optimal Lagrange multiplier (OLM)
constraint [19, 20].  The OLM constraint removes the
self-referentiality caused by the T-M-P constraints by introducing
centered mean values.

A recent formulation by Ferri, Martinez, and Plastino [21] has
demonstrated a methodology to ``rescue" the linear constraints in
maximum (Tsallis) entropy models, and, has related solutions
obtained using the linear, C-T, and, T-M-P constraints. This
formulation [21] has commonality with the studies of Wada and
Scarfone [22], Bashkirov [23], and, Di Sisto \textit{et. al.} [24].
This paper extends the work in [16], by employing the
Ferri-Martinez-Plastino formulation [21] to formulate
self-consistent nonextensive RD models for $ 0 < q < 1 $ and $ q > 1
$.

Tsallis statistics is described by two separate ranges of the
nonextensivity parameter, i.e. $ 0 < q < 1 $ and $ q > 1 $. Within
the context of coding theory and learning theory, each range of $ q
$ has its own specific utility. Un-normalized Tsallis entropies take
different forms for $ 0 < q < 1 $ and $ q
> 1 $, respectively. For example, as defined in Section 2, for $ 0 <
q < 1 $, the generalized mutual entropy is of the form $ I_{0<q<1}
\left( {X;\tilde X} \right)  =  - \sum\limits_{x,\tilde x} {p\left(
{x,\tilde x} \right)\ln _q } \left( {\frac{{p\left( x \right)p\left(
{\tilde x} \right)}}{{p\left( {x,\tilde x} \right)}}} \right) $.

For $ q > 1 $, as described in Section 2, the generalized mutual
entropy is defined by $ I_{q > 1}(X;\tilde X)=S_q(X)+S_q(\tilde
X)-S_q(X,\tilde X) $, where $ S_q(X) $ and $ S_q(\tilde X) $ are the
marginal Tsallis entropies for the random variables $ X $ and $
\tilde X $, and,  $ S_q(X,\tilde X)$ is the joint Tsallis entropy.
Unlike the Boltzmann-Gibbs-Shannon case, $ I_{0<q<1}(X;\tilde X) $
can never acquire the form of $ I_{q > 1} $, and vice versa. While
the form of $ I_{0<q<1}(X;\tilde X) $ is important in a number of
applications of practical interest in coding theory and learning
theory, un-normalized Tsallis entropies for $ q>1 $ demonstrate a
number of important properties such as the \textit{generalized data
processing inequality} and the \textit{generalized Fano inequality}
[12].

It may be noted that normalized Tsallis entropies do exhibit the
\textit{generalized data processing inequality} and the
\textit{generalized Fano inequality} [11].  As pointed out by Abe
[25], normalized Tsallis entropies do not possess Lesche stability.
However, for applications in communications theory and learning
theory, the local stability criterion of Yamano [26] may be evoked
to justify the use of normalized Tsallis entropies described in
terms of \textit{escort probabilities}.  Ongoing studies, which will
be reported elsewhere, have established the relation between the
solutions of generalized RD theory for un-normalized Tsallis
entropies using linear constraints that are reported in this paper,
and, normalized Tsallis entropies using T-M-P constraints defined in
terms of \textit{escort probabilities}, in a manner similar to that
employed by Wada and Scarfone [27].

To reconcile the different forms of the generalized mutual entropy
for $ 0<q<1 $ and $ q>1 $, the \textit{additive duality of
nonextensive statistics} [28] is evoked in Section  3.   This
results in dual Tsallis entropies characterized by
re-parameterization of the nonextensivity parameter $ q^* = 2-q $,
results in a \textit{dual generalized RD theory}.  An important
feature of dual Tsallis entropies is the similarity of the forms of
the Tsallis entropies with their counterparts in
Boltzmann-Gibbs-Shannon statistics, the difference being $
\log(\bullet) \rightarrow \ln_{q^*}(\bullet) $ [27].

In this paper, Tsallis entropies characterized by a nonextensivity
parameter $ q $ are called \textit{q-Tsallis entropies}. Similarly,
those characterized by the re-parameterized nonextensivity parameter
$ q^* $ are called \textit{$ q^* $-Tsallis entropies}. The two forms
of Tsallis entropies may be used in conjunction to obtain a
self-consistent description of nonextensive phenomena [29].

Summing up, this Section outlines the material presented in this
paper. The basic theory of \textit{$ q $-Tsallis entropies}, and,
\textit{$ q^* $-Tsallis entropies} is described in Sections 2 and 3,
respectively.  Section 3 also derives select information theoretic
properties for \textit{$ q^* $-Tsallis entropies}. Section 4 defines
the generalized statistics RD problem, and, describes alternating
minimization numerical algorithms within the ambit of nonextensive
statistics. The mathematical basis underlying nonextensive
alternating minimization algorithms (rate distortion), and
subsequently, alternating maximization algorithms (channel
capacity), is also derived in Section 4. This is accomplished in
Lemma 1 of this paper, by extending the positivity conditions in
Lemma 13.8.1 in [3] to the case of Tsallis statistics.  Section 5
extends prior studies [16] by deriving variational principles for
both, a generalized RD theory, and, a dual generalized RD theory.
The practical implementation of a nonextensive alternating
minimization algorithm is also described in Section 5.

Section 6 presents numerical simulations that demonstrate the
efficacy of the generalized RD theory \textit{vis-\'{a}-vis}
equivalent formulations derived within the Boltzmann-Gibbs-Shannon
framework. It is demonstrated that the generalized RD theory
possesses a lower threshold for the \textit{compression
information}, as compared with equivalent extensive
Boltzmann-Gibbs-Shannon RD models. \textit{This feature has immense
potential significance in data compression applications}. Section 7
concludes this paper by summarizing salient results, and, briefly
highlighting qualitative extensions that will be presented in
forthcoming publications.

\section{Tsallis entropies}

By definition, the un-normalized Tsallis entropy, is defined in
terms of discrete variables as [1, 2]
\begin{equation}
\begin{array}{l}
S_q \left( X \right) = -\frac{{1 - \sum\limits_x {p^q \left( x
\right)} }}{{1 - q}}; \sum\limits_x {p\left( x \right)}  = 1. \\
\end{array}
\end{equation}
The constant $ q $ is referred to as the nonextensivity parameter.
Given two independent variables $ X $ and $ Y $, one of the
fundamental consequences of nonextensivity is demonstrated by the
\textit{pseudo-additivity} relation
\begin{equation}
S_q \left( {XY} \right) = S_q \left( X \right) + S_q \left( Y
\right) + \left( {1 - q} \right)S_q \left( X \right)S_q \left( Y
\right).
\end{equation}
Here, (1) and (2) imply that extensive statistics is recovered as $
q \to 1 $.  Taking the limit $ q \to 1 $ in (1) and evoking
l'Hospital's rule,  $ S_q \left( X \right) \to S\left( X \right) $,
i.e.,  the Shannon entropy. The generalized Kullback-Leibler
divergence (K-Ld) is of the form [30, 31]
\begin{equation}
D_{K-L}^q \left[ {p\left( X \right)\left\| {r\left( X \right)}
\right.} \right] = \sum\limits_x {p\left( x \right)} \frac{{\left(
{\frac{{p\left( x \right)}}{{r\left( x \right)}}} \right)^{q - 1}  -
1}}{{q - 1}}. \
\end{equation}
Akin to the Tsallis entropy, the generalized K-Ld obeys the
\textit{pseudo-additivity} relation [31].  Nonextensive statistics
is intimately related to \textit{q-deformed }algebra and calculus
(see [32] and the references within). The \textit{q-deformed}
logarithm and exponential are defined as [32]
\begin{equation}
\begin{array}{l}
ln _q \left( x \right) = \frac{{x^{1 - q} - 1}}{{1 - q}}, \\
and, \\
exp_q(x) = \left\{ \begin{array}{l}
 \left[ {1 + \left( {1 - q} \right)x} \right]^{^{{\raise0.7ex\hbox{$1$} \!\mathord{\left/
 {\vphantom {1 {\left( {1 - q} \right)}}}\right.\kern-\nulldelimiterspace}
\!\lower0.7ex\hbox{${\left( {1 - q} \right)}$}}} } ;1 + \left( {1 - q} \right)x \ge 0 \\
 0;otherwise, \\
\end{array} \right.
\end{array}
\end{equation}
respectively. Before proceeding further, three important relations
from \textit{q-deformed} algebra, employed in this paper, are stated
[12, 32]
\begin{equation}
\begin{array}{l}
 \ln _q \left( {\frac{x}{y}} \right) = y^{q - 1} \left( {\ln _q x - \ln _q y} \right), \\
 \ln _q \left( {xy} \right) = \ln _q x + x^{1 - q} \ln _q y, \\
 and, \\
 \ln _q \left( {\frac{1}{x}} \right) =  - x^{q - 1} \ln _q x. \\
 \end{array}
\end{equation}
The un-normalized Tsallis entropy (1), conditional Tsallis entropy,
joint Tsallis entropy , and, the generalized K-Ld (3) may thus be
written as [12, 30, 31]
\begin{equation}
\begin{array}{l}
S_q \left( X \right) =  - \sum\limits_x {p\left( x
\right)} ^q \ln _q p\left( x \right), \\
 S_q \left( {\left. {\tilde X} \right|X} \right) =  - \sum\limits_x {\sum\limits_{\tilde x} {p\left( {x,\tilde x} \right)^q \ln _q p\left( {\left. {\tilde x} \right|x} \right)} } , \\
 S_q \left( {X,\tilde X} \right) =  - \sum\limits_x {\sum\limits_{\tilde x} {p\left( {x,\tilde x} \right)^q \ln _q p\left( {x,\tilde x} \right)} }  \\
=S_q(X)+S_q(\tilde X|X)=S_q(\tilde X)+ S_q(X|\tilde X),\\
and,\\
D_{K-L}^q \left[ {p\left( X \right)\left\| {r(X)} \right.} \right] =
- \sum\limits_x {p\left( x \right)}  \ln _q \frac{{r(x)}}{{p(x)}},
\end{array}
\end{equation}
respectively.  The joint convexity of the generalized K-Ld for $ q >
0 $ is established by the relation [33, 34]
\begin{equation}
\begin{array}{l}
D_{K - L}^q \left[ {\left. {\sum\limits_\alpha  {\eta _\alpha  p_a }
} \right\|\sum\limits_\alpha  {\eta _\alpha  r_a } } \right] \le
\sum\limits_\alpha  {\eta _\alpha  } D_{K - L}^q \left[ {\left. {p_a
} \right\|r_a } \right], \\
\eta _\alpha   > 0,and,\sum\limits_\alpha {\eta _\alpha  }  =
1. \\
\end{array}
\end{equation}

In the framework of Boltzmann-Gibbs-Shannon statistics, the mutual
information may be expressed as [3] $ I \left( {X;\tilde X} \right)
= S \left( X \right) - S \left( {\left. X \right|\tilde X} \right) =
S \left( {\tilde X} \right) - S \left( {\left. {\tilde X} \right|X}
\right) =I(\tilde X;X) $.  This is the manifestation of the symmetry
of the mutual information within the Boltzmann-Gibbs-Shannon model.
Within the framework of nonextensive statistics, the inequalities
(\textit{sub-additivities})
\begin{equation}
S_q \left( {\left. X \right|\tilde X} \right) \le S_q \left( X
\right),and,S_q \left( {\left. {\tilde X} \right|X} \right) \le S_q
\left( {\tilde X} \right),
\end{equation}
do not generally hold true for $ 0 <q <1 $. \textit{Note that (8) is
only valid for $ q> 1 $ as noted by Furuichi} [12].  The
\textit{sub-additivities} (8) are required to hold true, for the
generalized mutual entropy to be defined by
\begin{equation}
\begin{array}{l}
 I_q \left( {X;\tilde X} \right) = S_q \left( X \right) - S_q \left( {\left. X \right|\tilde X} \right) = S_q \left( {\tilde X} \right) - S_q \left( {\left. {\tilde X} \right|X} \right) \\
  = S_q \left( X \right) + S_q \left( {\tilde X} \right) - S_q \left( {X,\tilde X} \right)=I_q(\tilde X;X); q > 1.\\
 \end{array}
\end{equation}
Note that Furuichi [12] has presented a thorough and exhaustive
qualitative extension of the analysis of Dar\'{o}czy [35], and has
proven that (9) holds true for un-normalized Tsallis entropies for $
q\geq1 $.  Further, for normalized Tsallis entropies, Yamano [11]
has elegantly established the symmetry described by (9) in the range
$ 0 < q <1 $.

The un-normalized generalized mutual entropy $ I_q(X;\tilde X) $, in
the range $ 0 <q <1 $, is defined by the generalized K-Ld between
the the joint probability $ p(X,\tilde  X) $ and the marginal
probabilities $ p(X) $ and $ p(\tilde X) $, respectively
\footnote{The $ \log(\bullet) $ in the extensive convex mutual
information [3] is replaced by $ \ln_q(\bullet) $ (4). Also refer to
Theorem 3 in this paper.}
\begin{equation}
\begin{array}{l}
 I_q \left( {X;\tilde X} \right) =  - \sum\limits_{x,\tilde x} {p\left( {x,\tilde x} \right)\ln _q } \left( {\frac{{p\left( x \right)p\left( {\tilde x} \right)}}{{p\left( {x,\tilde x} \right)}}} \right) \\
  = D_{K - L}^q \left[ {p\left( {X,\tilde X} \right)\left\| {p\left( X \right)p\left( {\tilde X} \right)} \right.} \right] \\
  = \left\langle {D_{K - L}^q \left[ {\left. {p\left( {\left. {\tilde
X} \right|X} \right)} \right\|p\left( {\tilde X} \right)} \right]}
\right\rangle _{p\left( x \right)}; 0<q<1. \\
\end{array}
\end{equation}

The \textit{sub-additivities} (8) do not generally hold true in the
range $ 0<q<1 $.  This forecloses the prospect of transparently
establishing the symmetry of the generalized mutual entropy in the
range $ 0<q<1$, in the manner akin to (9), and, the
Boltzmann-Gibbs-Shannon model.

From (10), the symmetry of the generalized mutual entropies $
I_q(X;\tilde X) $ and $ I_q(\tilde X;X) $ may be summarized as
\begin{equation}
\begin{array}{l}
 I_q \left( {X;\tilde X} \right) = D_{K - L}^q \left[ {p\left( {X,\tilde X} \right)\left\| {p\left( X \right)p\left( {\tilde X} \right)} \right.} \right]   \\
 = \left\langle {D_{K - L}^q \left[ {\left. {p\left( {\left. {\tilde x} \right|x} \right)} \right\|p\left( {\tilde x} \right)} \right]} \right\rangle _{p\left( x \right)}  = \left\langle {D_{K - L}^q \left[ {\left. {p\left( {\left. x \right|\tilde x} \right)} \right\|p\left( x \right)} \right]} \right\rangle _{p\left( {\tilde x} \right)} \\
 =  D_{K - L}^q \left[ {p\left( {\tilde X,X} \right)\left\| {p\left( \tilde X \right)p\left( {X} \right)} \right.} \right] =I_q \left( {\tilde X;X} \right).   \\
\end{array}
\end{equation}
It is important to note that the generalized K-Ld (6) $
D^q_{K-L}[p(X)||r(X)] $ is not symmetric. However, as described in
(11), if $ X $ and $ \tilde X $ are discrete random variables with
marginal's $ p(X) $ and $ p(\tilde X) $ and joint distribution $
p(X,\tilde X) $, then the generalized mutual entropy for $ 0<q<1 $
defined in terms of the generalized K-Ld is symmetric.  This is a
consequence of the Kolmogorov theorem which states that joint
distributions are always invariant to the ordering of the random
variables [36]. Specifically, the generalized K-Ld between $
p(X,\tilde X) $ and $ p(X)p(\tilde X) $ is identical to the
generalized K-Ld between $ p(\tilde X,X) $ and $ p(\tilde X)p(X) $.
The symmetry between the two distinct forms of generalized mutual
entropy for $ 0 < q < 1 $ and $ q^*>1 $, in a manner analogous to
(9) and the Botlzmann-Gibbs-Shannon case, may be proven with the aid
of the \textit{additive duality}.  The proof for this symmetry
relation is identical to the derivation of Theorem 3 in Section 3 of
this paper, and may be obtained on interchanging the nonextensivity
parameters $ q $ and $ q^*=2-q $, wherever they occur in (19).

\section{Dual Tsallis information theoretic measures}

This paper makes prominent use of the \textit{additive duality} in
nonextensive statistics. Setting $ q^*=2-q $, from (4) the
\textit{dual deformed} logarithm and exponential are defined as
\begin{equation}
\begin{array}{l}
 \ln _{q^*}  \left( x \right) =  - \ln _q  \left( {\frac{1}{x}} \right), and, \exp _{q^*}  \left( x \right) = \frac{1}{{\exp _q  \left( { - x} \right)}}. \\
 \end{array}
\end{equation}
The reader is referred to Naudts [29] for further details.

A dual Tsallis entropy defined by
\begin{equation}
S_{q^*} \left( X \right) =  - \sum\limits_x {p\left( x \right)\ln
_{q^*} } p\left( x \right),
\end{equation}
has already been studied in a maximum (Tsallis) entropy setting (for
example, see Wada and Scarfone [27]). It is important to note that
the $ q^* =2-q $ duality has been studied within the
Sharma-Taneja-Mittal framework by Kanniadakis, \textit{et. al.}
[37]. The following properties, however, have yet to be proven for
dual Tsallis entropies:  $ (a) $ the validity of (9) for dual
Tsallis entropies, and, $ (b) $ the adherence of the
dual Tsallis entropies to the chain rule [3, 12]. The task is undertaken below. \\

\textbf{Theorem 1}:  \textit{The dual Tsallis joint entropy obeys
the relation
\begin{equation}
\begin{array}{l}
 S_{q^*} \left( {X,\tilde X} \right) = S_{q^*} \left( X \right) + S_{q^*} \left( {\left. {\tilde X} \right|X} \right). \\
 \end{array}
\end{equation}}
Note that $  q^{*}=2-q $,  $ \ln_{q^*}x = \frac{{x^{1 - q^*} -
1}}{{1 -
q^*}} $. \\

\textbf{Proof}:  From (5) and (6)
\begin{equation}
\begin{array}{l}
 S_q \left( {X,\tilde X} \right) =  - \sum\limits_x {\sum\limits_{\tilde x} {p\left( {x,\tilde x} \right)^q } } \ln _q p\left( {x,\tilde x} \right) \\
  =  - \sum\limits_x {\sum\limits_{\tilde x} {p\left( {x,\tilde x} \right)^q } } \ln _q \left( {p\left( x \right)p\left( {\left. {\tilde x} \right|x} \right)} \right) \\
  =  - \sum\limits_x {\sum\limits_{\tilde x} {p\left( {x,\tilde x} \right)^q } } \left[ {\ln _q p\left( x \right) + p\left( x \right)^{1 - q} \ln _q p\left( {\left. {\tilde x} \right|x} \right)} \right] \\
  = \sum\limits_x {p\left( {x,\tilde x} \right)\ln _q \left( {\frac{1}{{p\left( x \right)}}} \right) + \sum\limits_x {p\left( x \right)} \sum\limits_{\tilde x} {p\left( {\left. {\tilde x} \right|x} \right)\ln _q \left( {\frac{1}{{p\left( {\left. {\tilde x} \right|x} \right)}}} \right)} }  \\
\Rightarrow S_{q^*} \left( {X,\tilde X} \right) =  - \sum\limits_x
{p\left( x \right)\ln _{q^*} p\left( x \right)}- \sum\limits_x
{\sum\limits_{\tilde x} {p\left( {x,\tilde x} \right)\ln _{q^ *  }
p\left( {\left. {\tilde x} \right|x} \right)} } \\
  = S_{q^*} \left( X \right) + S_{q^*} \left( {\left. {\tilde X} \right|X} \right). \\
\end{array}
\end{equation}
\textit{In conclusion, the dual Tsallis entropies acquire a form
identical to the Boltzmann-Gibbs-Shannon entropies, with $
\ln_{q^*}(\bullet) $ replacing $ \log(\bullet) $}.

\textbf{Theorem 2}:  \textit{Let $ X_1 ,X_2 ,X_3 ,...,X_n  $ be
random variables obeying the probability distribution $ p\left( {x_1
,x_2 ,x_3 ,...,x_n } \right) $, then  we have the chain rule
\begin{equation}
\begin{array}{l}
S_{q^*} \left( {X_1 ,X_2 ,X_3 ,...,X_n } \right) = \sum\limits_{i =
1}^n {S_{q^*} } \left( {\left. {X_i } \right|X_{i - 1} ,...,X_1 }
 \right). \\
\end{array}
\end{equation}}

\textbf{Proof}:  Theorem 2 is proved by induction on $ n $. Assuming
(16) holds true for some $ n $, (15) yields
\begin{equation}
\begin{array}{l}
 S_{q^*} \left( {X_1 ,X_2 ,X_3 ,...,X_{n + 1} } \right)  \\
 = S_{q^*} \left( {X_1 ,X_2 ,X_3 ,...,X_n } \right) + S_{q^*} \left( {\left. {X_{n + 1} } \right|X_n ,...,X_1 } \right) \\
  = \sum\limits_{i = 1}^n {S_{q^*} \left( {\left. {X_i } \right|X_{i - 1} ,...,X_1 } \right)}  + S_{q^*} \left( {\left. {X_{n + 1} } \right|X_n ,...,X_1 } \right), \\
 \end{array}
\end{equation}
which implies that (16) holds true for $ n+1 $. Theorem 2 implies
that the dual Tsallis entropies can support a parametrically
extended information theory.

\textbf{Theorem 3:  }The dual convex generalized mutual entropy is
described by\footnote{Here "$ \rightarrow $" denotes a
re-parameterization of the nonextensivity parameter, and, is not a
limit.}

\begin{equation}
\begin{array}{l}
I_{q^*}(X;\tilde X) =  - \sum\limits_x {\sum\limits_{\tilde x} {p\left( {x,\tilde x} \right)\ln _{q^ *  } } } \left( {\frac{{p\left( x \right)p\left( {\tilde x} \right)}}{{p\left( {x,\tilde x} \right)}}} \right) \\
\mathop   = \limits^{\left( q^* \rightarrow q \right)}S_q(X)+S_q(\tilde X)-S_q(X,\tilde X)=I_q(X;\tilde X);0<q^*<1,and,q>1. \\
\end{array}
\end{equation}

\textit{Note that in this paper, $ q^* \rightarrow q $ denotes
re-parameterization of the nonextensivity parameter defining the
information theoretic quantity from $ q^* $ to $ q $ by setting $
q^{*}=2-q $.  Likewise, $ q \rightarrow q^* $ denotes
re-parameterization from $ q $ to $ q^* $ by setting $ q=2-q^* $}.

\textbf{Proof: }

\begin{equation}
\begin{array}{l}
 I_{q^ *  } \left( {X;\tilde X} \right) =
 - \sum\limits_x {\sum\limits_{\tilde x} {p\left( {x,\tilde x} \right)\ln _{q^ *  } } } \left( {\frac{{p\left( x \right)p\left( {\tilde x} \right)}}{{p\left( {x,\tilde x} \right)}}} \right)=- \sum\limits_x {\sum\limits_{\tilde x} {p\left( {x,\tilde x} \right)\ln _{q^ *  } \left( {\frac{{p\left( {\tilde x} \right)}}{{p\left( {\left. {\tilde x} \right|x} \right)}}} \right)} }  \\
 \mathop  = \limits^{\left( a \right)}  - \sum\limits_x {\sum\limits_{\tilde x} {p\left( {x,\tilde x} \right)} } \left( {\ln _{q^ *  } p\left( {\tilde x} \right) + p\left( {\tilde x} \right)^{\left( {1 - q^ *  } \right)} \ln _{q^ *  } \left( {\frac{1}{{p\left( {\left. {\tilde x} \right|x} \right)}}} \right)} \right) \\
 =  - \sum\limits_{\tilde x} {p\left( {x,\tilde x} \right)\frac{{p\left( {\tilde x} \right)^{q^ *   - 1} }}{{p\left( {\tilde x} \right)^{q^ *   - 1} }}} \frac{{p\left( {\tilde x} \right)^{1 - q^ *  }  - 1}}{{1 - q^ *  }}   \\
  - \sum\limits_x {\sum\limits_{\tilde x} {p\left( x \right)p\left( {\left. {\tilde x} \right|x} \right)} } p\left( x \right)^{\left( {1 - q^ *  } \right)} p\left( {\left. {\tilde x} \right|x} \right)^{\left( {1 - q^ *  } \right)} \ln _{q^ *  } \left( {\frac{1}{{p\left( {\left. {\tilde x} \right|x} \right)}}} \right) \\
 \mathop  = \limits^{\left( b \right)}    - \sum\limits_{\tilde x} {p\left( {\tilde x} \right)^q } \frac{{p\left( {\tilde x} \right)^{1 - q}  - 1}}{{1 - q  }} + \sum\limits_x {\sum\limits_{\tilde x} {p\left( x \right)^q p\left( {\left. {\tilde x} \right|x} \right)^q } } \ln _q p\left( {\left. {\tilde x} \right|x} \right) \\
 \Rightarrow I_{q^*}(X;\tilde X)\mathop=\limits^{q^ * \to q  } S_q \left( {\tilde X} \right) - S_q \left( {\left. {\tilde X} \right|X} \right) \\
 \mathop  = \limits^{\left( c \right)}  S_q(X)+S_q(\tilde X)-S_q(X,\tilde X)
\mathop  = \limits^{q \to q^ *  } I_{q^ *  } \left( {\tilde X;X}
\right). \\
 \end{array}
\end{equation}
Here, $ (a) $ follows from (5), $ (b) $ follows from setting $
q^*=2-q $, and, $ (c) $ follows from (9).  Theorem 3 acquires a
certain significance especially when it may be proven [11] that the
convex form of the generalized mutual entropy (10) can never be
expressed in the form of Tsallis entropies (9).

Theorem 3 demonstrates that such a relation is indeed possible by
commencing with \textit{$ q^* $-Tsallis mutual entropy} and
performing manipulations that scale the generalized mutual entropy
from $ q^* $-space to $ q $-space, yielding a form akin to (9).
Interchanging the range of values and the connotations of $ q $ and
$ q^* $ respectively, such that $ 0 < q < 1 $ ( $ q^*
>1 $), Theorem 3 may be modified to justify defining the convex \textit{$ q
$-Tsallis mutual entropy} by (10).

\section{Nonextensive rate distortion theory and alternating
minimization schemes}
\subsection{Overview of rate distortion theory}

For a thorough exposition of RD theory, the interested reader is
referred to Section 13 in [3].  Let $ X $ be a discrete random
variables with a finite set of possible values $ \mathcal{X} $,
distributed according to $ p(x) $. Here, $ X $ is the \textit{source
alphabet}. Let $ \tilde X $ denote the \textit{reproduction
alphabet} (a compressed representation of $ X $).  Let, $
X=\{X_1,...,X_n\} $ and $ \tilde X= \{\tilde X_1,...,\tilde X_m\} $,
$ m<n $.  By definition, the partitioning of $ X $ dictates the
manner in which each element of the \textit{source alphabet} $ X $
relates to each element of the \textit{compressed representation} $
\tilde X $.  In RD theory, the partitioning is dictated by the
normalized transition probability $ p(\tilde x|x) $. When the
assignment of the elements of $ X \rightarrow \tilde X $ is
probabilistic, then the partitioning is referred to as \textit{soft
partitioning}.  When the assignment is deterministic, it is referred
to as \textit{hard partitioning}.  Initially, RD undergoes a soft
partitioning, and then as the process progresses, hard partitioning
occurs.

A standard measure that defines the quality of compression is the
rate of a \textit{code} with respect to a channel transmitting
between $ X $ and $ \tilde X $.  In generalized statistics, this
quantity is the generalized mutual entropy $ I_q(X;\tilde X) $
\footnote{Note that these arguments are adapted from the
Boltzmann-Gibbs-Shannon RD model, where $ q=1 $.}.  The quantity $
I_q(X;\tilde X) $ is defined as the\textit{ compression
information}, which is evaluated on the basis of the joint
probability $ p(x)p(\tilde x|x) $.  Low values of $ I_q(X;\tilde X)
$ imply a more compact representation, and thus better compression.
An extreme case would be where $ \tilde X $ has only one element
(cardinality of $ \tilde\mathcal{X}=1 $, $ |\tilde\mathcal{X}|=1 $),
resulting in $ I_q(X;\tilde X) =0 $.

The physics underlying RD remains unchanged regardless of the range
of $ q $.  However, the generalized mutual entropy possesses
different properties for nonextensivity parameters in i) $ 0 <q< 1$
and ii) $ q>1 $ respectively.  For example, information theoretic
and physics based features intrinsic to RD theory, that may not be
comprehensively described using one range of nonextensivity
parameter, may be analyzed with greater coherence using the \textit{additive
duality}.  To obtain a deeper insight into the generic process of
nonextensive RD, the case of $ q>1 $ is briefly examined.

\textit{This is a simple example where nonextensive models, having
nonextensivity parameters in the ranges i) $ 0<q<1 $ and ii) $ q>1
$, may be employed to complement each other. The models may help to
understand the physics of a certain  problems
 by employing the additive duality. }  In this case, the two
 different ranges of $ q $ are employed with the aid of the \textit{additive
 duality}, to qualitatively and quantitatively describe the extreme limits (scenarios) of the generalized RD model, using a
 \textit{sender-receiver} description.  These extreme limits are: $
 i) $ no communication between $ \mathcal{X} $ and $  \tilde\mathcal{X}  $, and, $ ii) $ perfect communication between $ \mathcal{X} $ and $ \tilde\mathcal{X}  $, respectively.

 Specifically, in Section 5.1 of this paper, the mathematical
 (quantitative) analysis of the generalized RD model is performed in
 the range $ 0<q<1 $ using the generalized mutual entropy described
 by (10). However, a deeper qualitative description of the process may be obtained by complementing the form of the
 generalized mutual entropy used in Section 5.1, with the form of the generalized mutual entropy valid in the range $ q>1 $ (described by (9)), with the aid of the \textit{additive duality}.  As stated in Sections 1 and 2, the generalized mutual entropy $
I_q(X;\tilde X) $ cannot be stated in the form described by (9) for
$ 0<q<1 $. Employing the definition of the generalized mutual
entropy for $ q > 1 $ (9), we obtain $ I_q(X;\tilde X) = S_q(\tilde
X)-S_q(\tilde X|X) $.  Let $ \mathcal{X} $ be inhabited by a sender
and $ \tilde\mathcal{X} $  by a receiver. Here, $ S_q(\tilde X) $ is
the prior uncertainty the receiver has about the sender's signal,
which is diminished by $ S_q(\tilde X|X) $ as the signal is
received.  The difference yields $ I_q(X;\tilde X) $.  As an
example, in case there is no communication at all, then $ S_q(\tilde
X|X)=S_q(\tilde X) $, and, $ I_q(X;\tilde X)=0 $.

Alternatively, if the communication channel is perfect and the
received signal $ \tilde X $  is identical to the signal $ X $ at
the sender (the \textit{source alphabet} is simply copied as the
\textit{reproduction alphabet}), then $ S_q(\tilde X|X)=0 $ and $
I_q(X;\tilde X)=S_q(\tilde X)=S_q(X) $. In Boltzmann-Gibbs-Shannon
statistics ($ q=1 $), this is called the Shannon upper bound [3]. In
nonextensive statistics, this quantity is hereafter referred to as
the \textit{Tsallis upper bound}.

The compression information may always be reduced by using only a
single value of $ \tilde X $, thereby ignoring the details in $ X $.
This requires an additional constraint called the \textit{distortion
measure}.  The distortion measure is denoted by $ d(x,\tilde x) $
and is taken to be the Euclidean square distance for most problems
in science and engineering [3,6].

Given $ d(x,\tilde x) $, the partitioning of $ X $ induced by $
p(\tilde x|x) $ has an expected distortion $ D=<d(x,\tilde
x)>_{p(x,\tilde x)} = \sum\limits_{x,\tilde x} {p\left( {x,\tilde x}
\right)} d \left( {x,\tilde x} \right) $ \footnote{Note that in this
paper, $ \left\langle \bullet \right\rangle_{p(\bullet)} $ denotes
the expectation with respect to the probability $ p(\bullet) $.}.
Note that $ D $ is mathematically equivalent to the internal energy
in statistical physics.

The RD function is [3, 4]
\begin{equation}
R_q \left( D \right) = \mathop {\min }\limits_{p\left( {\left.
{\tilde x} \right|x} \right):\left\langle {d\left( {x,\tilde x}
\right)} \right\rangle _{p\left( {x,\tilde x} \right)}  \le D} I_q
\left( {X;\tilde X} \right), \
\end{equation}

where, $ R_q(D) $ is the minimum of the \textit{compression
information}.   This implies that $ R_q(D) $ is the minimum
achievable nonextensive \textit{compression information}, where the
minimization is carried out over all normalized transition
probabilities $ p(\tilde x|x) $, for which the distortion constraint
is satisfied.  As depicted in Fig. 1, $ R_q(D) $ is a non-increasing
convex function of D in the \textit{distortion-compression plane}.
The $ R_q(D) $ function separates the \textit{distortion-compression
plane} into two regions. The region above the curve is known as the
\textit{rate distortion region}, and, corresponds to all achievable
\textit{distortion-compression pairs} $ \{D;I_q(X;\tilde X) \} $.

On the other hand, the region below the curve is known as the
\textit{non-achievable region}, where compression cannot
occur.\textit{ The major feature of nonextensive RD models is that
the RD curves inhabit the non-achievable region of those obtained
from Boltzmann-Gibbs-Shannon statistics}.  This implies that
nonextensive RD models can perform data compression in regimes not
achievable by equivalent extensive RD models.

Obtaining $ R_q(D) $ involves minimization of the nonextensive RD
Lagrangian (free energy) [4]
\begin{equation}\label{1}
L_{RD}^q \left[ p({\tilde x\left| x \right.} )\right] \
 = I_q \left( {X;\tilde X} \right) + \tilde\beta \left\langle
{d\left( {x,\tilde x} \right)} \right\rangle _{p\left( {x,\tilde x}
\right)},
\end{equation}
subject to the normalization of the conditional probability. Here, $
\tilde\beta=q\beta $ (see Section 5.1) is a Lagrange multiplier
called the \textit{nonextensive trade-off parameter}, where, $ \beta
$ is the \textit{inverse temperature} in Boltzmann-Gibbs-Shannon
statistics [5, 6].  Note that $
\tilde\beta $ is the nonextensive mathematical equivalent of the
Boltzmann-Gibbs-Shannon \textit{inverse temperature} $ \beta $.  It is important to note that $ \tilde\beta $ and $ \beta $ have different physical connotations.  The definition $ \tilde\beta=q\beta $
is specific to this paper, and is chosen to facilitate comparison between the extensive and nonextensive RD models.  More complex forms of the
\textit{nonextensive trade-off parameter} have been investigated
into. The results of this study will be reported elsewhere.

Here, (21) implies that RD theory is a trade-off between the
\textit{compression information} and the expected distortion. Taking
the variation of (21) over all normalized distributions $ p(\tilde
x|x) $, yields
\begin{equation}
\begin{array}{l}
 \delta L_{RD}^q \left[ p({\tilde x\left| x \right.} )\right] \ = \delta I_q \left( {X;\tilde X} \right) + \tilde\beta \delta \left\langle {d\left( {x,\tilde x} \right)} \right\rangle _{p\left( {x,\tilde x} \right)}  = 0 \\
\Rightarrow \frac{{\delta I_q \left( {X;\tilde X} \right)}}{{\delta \left\langle {d\left( {x,\tilde x} \right)} \right\rangle _{p\left( {x,\tilde x} \right)} }} =  - \tilde\beta,  \\
 \end{array}
\end{equation}
Here, (22) implies that the rate of change of the generalized mutual
entropy with respect to the expected distortion is
 called the RD curve, and a tangent drawn at any point on the RD
curve has a slope $ -\tilde\beta $.  To prove that the conditional
distribution $ p(\tilde x|x) $ represents a stationary point of $
L_{RD}^q[p(\tilde x|x)] $, (21) is subjected to a variational
minimization contingent to the normalization of $ p(\tilde x|x) $.
This procedure is detailed in Section 5 of this paper, for both
\textit{$ q $ and $ q^* $ Tsallis mutual entropies}.

\subsection{The nonextensive alternating minimization scheme}

The basis for alternating  minimization algorithm  (a class of
algorithms that include the Blahut-Arimoto scheme) \footnote{See
Chapter 13 in [3].} is to find the minimum distance between two
convex sets $ A $ and $ B $ in $ \mathcal{R}^n $ \footnote{Note that
the nonextensive alternating minimization algorithm presented herein
is not referred to as the nonextensive Blahut-Arimoto algorithm,
despite being an obvious extension, because the Blahut-Arimoto
scheme is synonymous with Boltzmann-Gibbs-Shannon statistics.}.
First, a point $ a \in A $ is chosen and a point $ b \in B $ closest
to it is found. This value of $ b $ is fixed and its closest point
in $ A $ is then found.  The above process is repeated till the
algorithm converges to a (global) minimum distance.

Extrapolating the Csisz\'{a}r-Tusn\'{a}dy theory [9] to the
nonextensive domain for two convex sets of probability
distributions, and considering the generalized mutual entropy (10)
as a distance measure between the joint probability $ p(x)p(\tilde
x|x) $, and the marginal probabilities $ p(x) $ and $ p(\tilde x) $
[3, 9], by extending the definition of the generalized K-Ld (3), the
nonextensive alternating minimization algorithm converges to the
minimum $ D^q_{K-L}[\bullet] $ between the two convex sets of
probability distributions ($ p(\tilde  x|x) $ and $ p(\tilde x) $,
respectively). It is important to note that the nonextensive
Pythagorean identity (triangular equality), which forms the basis of
any extension of the Csisz\'{a}r-Tusn\'{a}dy theory to the
nonextensive regime, has been established by Dukkipati \textit{et.
al.} [38].

Before proceeding any further, it is judicious to state the
leitmotif of this Section.  The procedure behind the alternating
minimization algorithm described herein assumes an \textit{a-priori}
minimization of the nonextensive RD Lagrangian (21), with respect to
conditional probabilities $ p(\tilde x|x) $,  for all normalized $
p(\tilde x|x) $, using the calculus of variations. This variational
minimization yields a canonical conditional probability $ p(\tilde
x|x) $.  The variational minimization procedure is presented in
Section 5.

Here, $ p(\tilde x|x) $ corresponds to the joint probability $
p(x,\tilde x)=p(x)p(\tilde x|x) $, which is employed to evaluate the
expected distortion $ D=<d(x,\tilde x)>_{p(x,\tilde x)} $.  It is
important to prove that, $ p(\tilde x) $ is a marginal probability
(or marginal) of $ p(x,\tilde x) $. This criterion ensures that
extremization with respect to $ p(\tilde x) $ further minimizes
(21).  Section 5.3 provides a discussion of the nonextensive
alternating minimization algorithm, from a practitioner's viewpoint.
Now, a result that establishes the positivity condition for nonextensive alternating
minimization schemes, central to RD
theory, is proven. Subsequently, this result is extended
to establish the positivity condition for nonextensive alternating maximization schemes, required for calculating the
channel capacity (Lemma 13.8.1 in [3]).

\textbf{Lemma 1}:  Let $ p(x)p(\tilde x|x) $ be a given joint
distribution. The prior distribution $ p(\tilde x) $ that minimizes
$ D^q_{K-L}[p(X)p(\tilde X|X)||p(X)p(\tilde X)] $ is the marginal
distribution to $ p^*(\tilde x) $ corresponding to $ p(\tilde x|x)
$, i.e.

\begin{equation}
\begin{array}{l}
 D_{K - L}^q \left[ {p\left( X \right)p\left( {\left. {\tilde X} \right|X} \right)\left\| {p\left( X \right)p^ *  \left( {\tilde X} \right)} \right.} \right] \\
= \mathop {\min }\limits_{p\left( {\tilde x} \right)} D_{K - L}^q \left[ {p\left( X \right)p\left( {\left. {\tilde X} \right|X} \right)\left\| {p\left( X \right)p\left( {\tilde X} \right)} \right.} \right], \\
 \end{array}
\end{equation}
where
\begin{equation}
p^ *  \left( {\tilde x} \right) = \sum\limits_x {p(x)p(\tilde x|x)}.
 \end{equation}

Also
\begin{equation}
\begin{array}{l}
- \sum\limits_{x,\tilde x} {p\left( x \right)p\left( {\left. {\tilde x} \right|x} \right)\ln _q \left( {\frac{{p\left( x \right)}}{{p^ *  \left( {\left. x \right|\tilde x} \right)}}} \right)} = -\mathop {\max }\limits_{p\left( {\left. x \right|\tilde x} \right)} \sum\limits_{x,\tilde x} {p\left( x \right)p\left( {\left. {\tilde x} \right|x} \right)\ln _q \left( {\frac{{p\left( x \right)}}{{p\left( {\left. x \right|\tilde x} \right)}}} \right)},  \\
 \end{array}
\end{equation}
where by the Bayes theorem
\begin{equation}
 p^ *  \left( {\left. x \right|\tilde x} \right) = \frac{{p\left( x \right)p\left( {\left. {\tilde x} \right|x} \right)}}{{\sum\limits_x {p\left( x \right)p\left( {\left. {\tilde x} \right|x} \right)} }}. \\
 \end{equation}
Note that (25) forms the basis for the computational implementation
of the channel capacity within the Tsallis statistics framework.

\textbf{Proof}:

\begin{equation}
\begin{array}{l}
 D_{K - L}^q \left[ {\left. {p\left( {X,\tilde X} \right)} \right\|p\left( X \right)p\left( {\tilde X} \right)} \right] - D_{K - L}^q \left[ {\left. {p\left( {X,\tilde X} \right)} \right\|p\left( X \right)p^ *  \left( {\tilde X} \right)} \right]   \\
  =  - \sum\limits_{x,\tilde x} {p\left( {x,\tilde x} \right)} \ln _q \left( {\frac{{p\left( {\tilde x} \right)p\left( x \right)}}{{p\left( {x,\tilde x} \right)}}} \right) + \sum\limits_{x,\tilde x} {p\left( {x,\tilde x} \right)} \ln _q \left( {\frac{{p^ *  \left( {\tilde x} \right)p\left( x \right)}}{{p\left( {x,\tilde x} \right)}}} \right) \\
  =  - \sum\limits_{x,\tilde x} {p\left( {x,\tilde x} \right)} \left[ {\ln _q \left( {\frac{{p\left( {\tilde x} \right)p\left( x \right)}}{{p\left( {x,\tilde x} \right)}}} \right) - \ln _q \left( {\frac{{p^ *  \left( {\tilde x} \right)p\left( x \right)}}{{p\left( {x,\tilde x} \right)}}} \right)} \right] \\
 \mathop  = \limits^{\left( a \right)}  - \sum\limits_{x,\tilde x} {p\left( {x,\tilde x} \right)} \frac{{\ln _q \left( {\frac{{p\left( {\tilde x} \right)p\left( x \right)}}{{p\left( {x,\tilde x} \right)}}} \right) - \ln _q \left( {\frac{{p^ *  \left( {\tilde x} \right)p\left( x \right)}}{{p\left( {x,\tilde x} \right)}}} \right)}}{{1 + \left( {1 - q} \right)\ln _q \left( {\frac{{p^ *  \left( {\tilde x} \right)p\left( x \right)}}{{p\left( {x,\tilde x} \right)}}} \right)}}\left[ {1 + \left( {1 - q} \right)\ln _q \left( {\frac{{p^ *  \left( {\tilde x} \right)p\left( x \right)}}{{p\left( {x,\tilde x} \right)}}} \right)} \right] \\
  =  - \sum\limits_{x,\tilde x} {p\left( {x,\tilde x} \right)} \ln _q \left( {\frac{{p\left( {\tilde x} \right)}}{{p^ *  \left( {\tilde x} \right)}}} \right)\left[ {1 + \left( {1 - q} \right)\ln _q \left( {\frac{{p^ *  \left( {\tilde x} \right)p\left( x \right)}}{{p\left( {x,\tilde x} \right)}}} \right)} \right] \\
 \mathop  = \limits^{\left( b \right)}  - \sum\limits_{x,\tilde x} {p\left( {x,\tilde x} \right)} \ln _q \left( {\frac{{p\left( {\tilde x} \right)}}{{p^ *  \left( {\tilde x} \right)}}} \right)\frac{{\left[ {\sum\limits_{x,\tilde x} {p\left( {x,\tilde x} \right)}  + \left( {1 - q} \right)\sum\limits_{x,\tilde x} {p\left( {x,\tilde x} \right)} \ln _q \left( {\frac{{p^ *  \left( {\tilde x} \right)p\left( x \right)}}{{p\left( {x,\tilde x} \right)}}} \right)} \right]}}{{\sum\limits_{x,\tilde x} {p\left( {x,\tilde x} \right)} }} \\
 =  - \sum\limits_{\tilde x} {\underbrace {\sum\limits_x {p\left( x \right)p\left( {\left. {\tilde x} \right|x} \right)} }_{p^ *  \left( {\tilde x} \right)}\ln _q \left( {\frac{{p\left( {\tilde x} \right)}}{{p^ *  \left( {\tilde x} \right)}}} \right)}\left[ {1 + \left( {q - 1} \right)\left( { - \sum\limits_{x,\tilde x} {p\left( {x,\tilde x} \right)\ln _q \left( {\frac{{p^ *  \left( {\tilde x} \right)p\left( x \right)}}{{p\left( {x,\tilde x} \right)}}} \right)} } \right)} \right]   \\
 \mathop  = \limits^{\left( c \right)} D_{K - L}^q \left[ {\left. {p^ *  \left( {\tilde X} \right)} \right\|p\left( {\tilde X} \right)} \right] \\
  \times \left[ {1 + \left( {q - 1} \right)D_{K - L}^q \left[ {\left. {p\left( {X,\tilde X} \right)} \right\|p\left( X \right)p^ *  \left( {\tilde X} \right)} \right]} \right] >
  0;\\
  \forall \left\{ \begin{array}{l}
 0 < q < 1, \\
 0 < D_{K - L}^q \left[ {p(X;\tilde X)\left\| {p(X)p^ *  (\tilde X)} \right.} \right] < 1, \\
 0 < D_{K - L}^q \left[ {p^ *  (\tilde X)\left\| {p(\tilde X)} \right.} \right] < 1. \\
 \end{array} \right.
\end{array}
\end{equation}

Here, $ (a) $ sets up the expression to employ the
\textit{q-deformed} algebra definition $ \frac{{\left[ {\ln _q a -
\ln _q b} \right]}}{{\left[ {1 + \left( {1 - q} \right)\ln _q b}
\right]}} = \ln _q \left( {\frac{a}{b}} \right) $, by multiplying
and dividing by $ \left[ {1 + \left( {1 - q} \right)\frac{{p^ *
\left( {\tilde x} \right)p\left( x \right)}}{{p\left( {x,\tilde x}
\right)}}} \right] $, $ (b) $ follows by multiplying and dividing $
\left[ {1 + \left( {1 - q} \right)\ln _q \left( {\frac{{p^ * \left(
{\tilde x} \right)p\left( x \right)}}{{p\left( {x,\tilde x}
\right)}}} \right)} \right] $ by $ p(x,\tilde x) $ and summing over
$ x $ and $ \tilde x $. Note that $ {\sum\limits_{x,\tilde x}
{p\left( {x,\tilde x} \right)} } =1 $, $ (c) $ establishes the
positivity condition and proves that

\begin{equation}
p^ *  \left( {\tilde x} \right) = \sum\limits_x {p(x)p(\tilde x|x)},  \\
\end{equation}

The second part of Lemma 1 is  proven as follows, based on the
validity of (24). From (24) and (26)
\begin{equation}
p\left( x \right)p\left( {\left. {\tilde x} \right|x} \right) = p(\tilde x)p^ *  \left( {\left. x \right|\tilde x} \right) \\
\end{equation}

Thus, the positivity condition is established as
\begin{equation}
\begin{array}{l}
 - \sum\limits_{x,\tilde x} {p\left( x \right)p\left( {\left. {\tilde x} \right|x} \right)\ln _q } \left( {\frac{{p\left( x \right)}}{{p^ *  \left( {\left. x \right|\tilde x} \right)}}} \right) + \mathop {\max }\limits_{p\left( {\left. x \right|\tilde x} \right)} \sum\limits_{x,\tilde x} {p\left( x \right)p\left( {\left. {\tilde x} \right|x} \right)\ln _q } \left( {\frac{{p\left( x \right)}}{{p\left( {\left. x \right|\tilde x} \right)}}}
 \right)\\
\mathop  =   - \sum\limits_{x,\tilde x} {p\left( x \right)p\left( {\left. {\tilde x} \right|x} \right)\left[ {\ln _q \left( {\frac{{p\left( x \right)}}{{p^ *  \left( {\left. x \right|\tilde x} \right)}}} \right) - \ln _q \left( {\frac{{p\left( x \right)}}{{p\left( {\left. x \right|\tilde x} \right)}}} \right)} \right]}  \\
 \mathop  = \limits^{\left( a \right)}  - \sum\limits_{x,\tilde x} {p\left( x \right)p\left( {\left. {\tilde x} \right|x} \right)\frac{{\ln _q \left( {\frac{{p\left( x \right)}}{{p^ *  \left( {\left. x \right|\tilde x} \right)}}} \right) - \ln _q \left( {\frac{{p\left( x \right)}}{{p\left( {\left. x \right|\tilde x} \right)}}} \right)}}{{1 + \left( {1 - q} \right)\ln _q \left( {\frac{{p\left( x \right)}}{{p\left( {\left. x \right|\tilde x} \right)}}} \right)}}\left[ {1 + \left( {1 - q} \right)\ln _q \left( {\frac{{p\left( x \right)}}{{p\left( {\left. x \right|\tilde x} \right)}}} \right)} \right]}  \\
 \mathop  =   - \sum\limits_{x,\tilde x} {p\left( x \right)p\left( {\left. {\tilde x} \right|x} \right)\ln _q \left( {\frac{{p\left( {\left. x \right|\tilde x} \right)}}{{p^ *  \left( {\left. x \right|\tilde x} \right)}}} \right)\left[ {1 + \left( {1 - q} \right)\ln _q \left( {\frac{{p\left( x \right)}}{{p\left( {\left. x \right|\tilde x} \right)}}} \right)} \right]}  \\
 \mathop  =  - \sum\limits_{x,\tilde x} {p\left( x \right)p\left( {\left. {\tilde x} \right|x} \right)\ln _q \left( {\frac{{p\left( {\left. x \right|\tilde x} \right)}}{{p^ *  \left( {\left. x \right|\tilde x} \right)}}} \right)\frac{{\left[ {\sum\limits_{x,\tilde x} {p\left( {x,\tilde x} \right)}  + \left( {1 - q} \right)\sum\limits_{x,\tilde x} {p\left( {x,\tilde x} \right)} \ln _q \left( {\frac{{p\left( x \right)}}{{p\left( {\left. x \right|\tilde x} \right)}}} \right)} \right]}}{{\sum\limits_{x,\tilde x} {p\left( {x,\tilde x} \right)} }}}  \\
 \mathop  = \limits^{\left( b \right)}  - \sum\limits_{x,\tilde x} {p\left( {\tilde x} \right)p^ *  \left( {\left. x \right|\tilde x} \right)\ln _q \left( {\frac{{p\left( {\left. x \right|\tilde x} \right)}}{{p^ *  \left( {\left. x \right|\tilde x} \right)}}} \right)\left[ {\sum\limits_{x,\tilde x} {p\left( {x,\tilde x} \right)}  + \left( {1 - q} \right)\sum\limits_{x,\tilde x} {p\left( {x,\tilde x} \right)} \ln _q \left( {\frac{{p\left( \tilde x \right)p\left( {x} \right)}}{{p\left( {\tilde x} \right)p\left( {\left. x \right|\tilde x} \right)}}} \right)} \right]}  \\
 \mathop
 = \sum\limits_{\tilde x} {p  \left( {\tilde x} \right)} \left( { - \sum\limits_x {p^ *  \left( {\left. x \right|\tilde x} \right)\ln _q \left( {\frac{{p\left( {\left. x \right|\tilde x} \right)}}{{p^ *  \left( {\left. x \right|\tilde x} \right)}}} \right)} } \right)\\
 \times\left[ {1 + \left( {q - 1} \right)\left( { - \sum\limits_{x,\tilde x} {p\left( {x,\tilde x} \right)\ln _q \left( {\frac{{p\left( {\tilde x} \right)p\left( x \right)}}{{p\left( {x,\tilde x} \right)}}} \right)} } \right)} \right] \\
\mathop  = \limits^{\left( c \right)}  \sum\limits_{\tilde x} {p\left( {\tilde x} \right)} D_{K - L}^q \left[ {\left. {p^ *  \left( {\left. X \right|\tilde X = \tilde x} \right)} \right\|p\left( {\left. X \right|\tilde X = \tilde x} \right)} \right] \\
  \times \left[ {1 + \left( {q - 1} \right)D_{K - L}^q \left[ {\left. {p\left( {X,\tilde X} \right)} \right\|p\left( X \right)p \left( {\tilde X} \right)} \right]} \right] > 0; \\
\forall \left\{ \begin{array}{l}
 0 < q < 1, \\
 0 < D_{K - L}^q \left[ {p(X;\tilde X)\left\| {p(X)p(\tilde X)} \right.} \right] < 1, \\
 0 < D_{K - L}^q \left[ {p^ *  (\left. X \right|\tilde X = \tilde x)\left\| {p(\left. X \right|\tilde X = \tilde x)} \right.} \right] < 1. \\
 \end{array} \right.
 \end{array}
\end{equation}

Here, $ (a) $ sets up the expression to employ the
\textit{q-deformed} algebra definition $ \frac{{\left[ {\ln _q a -
\ln _q b} \right]}}{{\left[ {1 + \left( {1 - q} \right)\ln _q b}
\right]}} = \ln _q \left( {\frac{a}{b}} \right) $, by multiplying
and dividing by $  \left[ {1 + \left( {1 - q} \right)\frac{{p\left(
x \right)}}{{p\left( {\left. x \right|\tilde x} \right)}}} \right]
$, $ (b) $ evokes (29) followed by multiplying and dividing $ \left[
{1 + \left( {1 - q} \right)\ln _q \left( {\frac{{p \left( {\tilde x}
\right)p\left( x \right)}}{{p\left( {x,\tilde x} \right)}}} \right)}
\right] $ by $ p(x,\tilde x) $ and summing over $ x $ and $ \tilde x
$. Note that $ {\sum\limits_{x,\tilde x} {p\left( {x,\tilde x}
\right)} } =1 $, $ (c) $ establishes the positivity condition.

\textit{Here, the two parts of Lemma 1 ((23) and (25)) provide a
generalized statistics framework for deriving both alternating
minimization and alternating maximization schemes, using the
celebrated Csisz\'{a}r-Tusn\'{a}dy theory.} Note that the results in
Boltzmann-Gibbs-Shannon statistics is obtained by setting $ q =1 $
(see Lemma 13.8.1 in [3]).

Lemma 1 is of great importance in deriving generalized statistics
extensions of the Expectation Maximization (EM) algorithm [39]. The
complete efficacy of Lemma 1 will be demonstrated in a future
publication, which studies the information bottleneck method [7]
within the framework of Tsallis statistics.

 \textit{The positivity conditions  (27) and (30) cannot be derived
without the aid of \textit{q-deformed} algebra [32]}. It is
important to note that the first part of Lemma 1 establishes the
fact that the marginal probability $ p^*(\tilde x) $, defined by
(24),  minimizes the generalized mutual entropy and hence the
generalized statistics RD Lagrangian (21), for the range $ 0<q<1 $.
Any other form of marginal probability other than (24), which may be
obtained without the use of q-deformed algebra, could have a
two-fold debilitating effect on the generalized statistics RD model
presented in this paper, and the results of Lemma 1. First, Bayes'
theorem (26) would be violated. Next, the possibility exists wherein
the RD Lagrangian (21) could be maximized instead of being
minimized, for certain values of $ q $.

Specifically, if the rules of \textit{q-algebra} are neglected, an
expression of the form:  $ p^*\left( {\tilde x} \right) =
\sum\limits_x {p\left( x \right)^{\alpha \left( q \right)} p\left(
{\left. {\tilde x} \right|x} \right)^{\beta \left( q \right)} } $,
where $ \alpha(q) $ and $ \beta(q) $ are some functions of the
nonextensivity parameter $ q $, is obtained.  Apart from violating
(26), such a form of $ p^*(\tilde x) $ could result in a
\textit{maximization} of (21) for certain values of $ q $, thereby
invalidating the nonextensive alternating minimization procedure.

Note that minimum $ D^q_{K-L}[p(X)p(\tilde X|X)||p(X)p(\tilde X)] $
is exactly the convex generalized mutual entropy $ I_q(X;\tilde X) $
calculated on the basis of the joint distribution $ p(x)p(\tilde
x|x) $. Thus, $ D^q_{K-L}[p(X)p(\tilde X|X)||p(X)p(\tilde X)] $ is
an upper bound for the \textit{compression information} term $
I_q(X;\tilde X) $, with equality achieved only when $ p(\tilde x) $
is set to the marginal distribution of $ p(x)p(\tilde x|x) $.  The
above proposition encourages the casting of the generalized RD
function as a double minimization

\begin{equation}
R_q \left( D \right) = \mathop {\min }\limits_{\left\{ {p\left(
{\tilde x} \right)} \right\}} \mathop {\min }\limits_{\left\{
{p(\tilde x|x):\left\langle {d\left( {x,\tilde x} \right)}
\right\rangle \le D} \right\}} D_{K - L}^q \left[ {\left. {p\left( X
\right)p\left( {\left. {\tilde X} \right|X} \right)} \right\|p\left(
X \right)p\left( {\tilde X} \right)} \right].
\end{equation}

 Given $ A $ a set of joint distributions $ p(x,\tilde x) $ with
 marginal $ p(x) $ that satisfy the distortion constraint, and, if $
 B $ is the set of product distributions $ p(\tilde x)p(x) $ with
 some normalized $ p(\tilde x) $, then

 \begin{equation}
 R_q \left( D \right) = \mathop {\min }\limits_{b \in B} \mathop {\min }\limits_{a \in A} D_{K - L}^q \left[ {\left. a \right\|b}\right].
 \end{equation}

Note that like the Blahut-Arimoto algorithm, the nonextensive
alternating minimization scheme does not possess a unique solution.
Extension of the above theory to the case of the \textit{dual} generalized
mutual entropy (parameterized by $ q^*=2-q $) is identical and straightforward.

\section{Nonextensive rate distortion variational principles}

This Section closely parallels the approach followed in Section 13.7 of [3].

\subsection{Case for $ 0<q<1 $}

\textbf{Lemma 2}: Variational minimization of the Lagrangian
\footnote{Valid for both discrete and continuous cases}
\begin{equation}
\begin{array}{l}
L_{RD}^q \left[ {p\left( {\left. {\tilde x} \right|x} \right)}
\right] = \sum\limits_x {\sum\limits_{\tilde x} {p\left( x
\right)p\left( {\left. {\tilde x} \right|x} \right)} } \frac{{\left(
{\frac{{p\left( {\left. {\tilde x} \right|x} \right)}}{{p\left(
{\tilde x} \right)}}} \right)^{q - 1}  - 1}}{{q - 1}}  \\
  + \tilde\beta \sum\limits_x {\sum\limits_{\tilde x} {d\left( {x,\tilde x} \right)p\left( x \right)} } p\left( {\left. {\tilde x} \right|x} \right) + \sum\limits_x {\lambda \left( x \right)\sum\limits_{\tilde x} {p\left( {\left. {\tilde x} \right|x} \right)}}, \\
\end{array}
\end{equation}
yields the canonical conditional probability $ p\left( {\tilde
x\left| x \right.} \right) = \frac{{p\left( {\tilde x} \right)\exp
_{q^ * }( -\beta^*(x) d\left( {x,\tilde x}) \right)}}{\tilde
Z(x,\beta^*(x)) } $,where, $\tilde \beta=q\beta $, $ \beta^* \left(
x \right) = \frac{\tilde\beta }{{(1-q)\tilde \lambda \left( x
\right)}} $, $ \tilde\lambda(x)=\frac{{\lambda \left( x
\right)}}{{p\left( x \right)}}$, and, $ \sum\limits_x
{\sum\limits_{\tilde x} {p\left( {x,\tilde x} \right)}}  = 1 $.

\textbf{Proof}: Taking the variational derivative of (33) for each $
x $ and $ \tilde x $, yields
\begin{equation}
\begin{array}{l}
 \frac{{\delta L_{RD}^q \left[ {p\left( {\left. {\tilde x} \right|x} \right)} \right]}}{{\delta p\left( {\left. {\tilde x} \right|x} \right)}} = 0 \\
  \Rightarrow \frac{\partial }{{\partial p\left( {\left. {\tilde x} \right|x}
\right)}}\left[ {\frac{{p\left( x \right)p\left( {\left. {\tilde x}
\right|x} \right)^q p\left( {\tilde x} \right)^{1 - q}  - 1}}{{q -
1}}} \right] +  \tilde\beta d\left( {x,\tilde x} \right)p\left( x \right) + \lambda \left( x \right) = 0 \\
 \mathop  \Rightarrow \limits^{} \frac{{qp\left( x \right)}}{{\left( {q - 1} \right)}}\left( {\frac{{p\left( {\left. {\tilde x} \right|x} \right)}}{{p\left( {\tilde x} \right)}}} \right)^{q - 1}  + \frac{{p\left( x \right)p\left( {\left. {\tilde x} \right|x} \right)^q }}{{\left( {q - 1} \right)}}\frac{{\partial p\left( {\tilde x} \right)^{1 - q} }}{{\partial p\left( {\left. {\tilde x} \right|x} \right)}} +  \tilde\beta d\left( {x,\tilde x} \right)p\left( x \right) + \lambda \left( x \right) = 0 \\
 \mathop  \Rightarrow \limits^{(a)} \frac{{qp\left( x \right)}}{{\left( {q - 1} \right)}}\left( {\frac{{p\left( {\left. {\tilde x} \right|x} \right)}}{{p\left( {\tilde x} \right)}}} \right)^{q - 1}  + \frac{{p\left( x \right)p\left( {\left. {\tilde x} \right|x} \right)^q }}{{\left( {q - 1} \right)}}\frac{\partial }{{\partial p\left( {\left. {\tilde x} \right|x} \right)}}\left( {\frac{{p\left( {\left. {\tilde x} \right|x} \right)^{1 - q} p\left( x \right)^{1 - q} }}{{p\left( {\left. x \right|\tilde x} \right)^{1 - q} }}} \right) \\
  +  \tilde\beta d\left( {x,\tilde x} \right)p\left( x \right) + \lambda \left( x \right) = 0 \\
 \mathop  \Rightarrow \limits^{} \frac{{qp\left( x \right)}}{{\left( {q - 1} \right)}}\left( {\frac{{p\left( {\left. {\tilde x} \right|x} \right)}}{{p\left( {\tilde x} \right)}}} \right)^{q - 1}  + \frac{{\left( {1 - q} \right)p\left( x \right)}}{{\left( {q - 1} \right)}}\left( {\frac{{p\left( x \right)}}{{p\left( {\left. x \right|\tilde x} \right)}}} \right)^{1 - q}  +  \tilde\beta d\left( {x,\tilde x} \right)p\left( x \right) + \lambda \left( x \right) = 0 \\
\mathop  \Rightarrow \limits^{} p\left( x \right)[ {\frac{1}{{\left( {q - 1} \right)}}\left( {\frac{{p\left( {\left. {\tilde x} \right|x} \right)}}{{p\left( {\tilde x} \right)}}} \right)^{q - 1}  +  \tilde\beta d\left( {x,\tilde x} \right) + \frac{{\lambda \left( x \right)}}{{p\left( x \right)}}}  ] = 0 \\
 \mathop  \Rightarrow \limits^{\left( b \right)} p\left( x \right)[ {\frac{1}{{\left( {q - 1} \right)}}\left( {\frac{{p\left( {\left. {\tilde x} \right|x} \right)}}{{p\left( {\tilde x} \right)}}} \right)^{q - 1}  + \tilde\beta d\left( {x,\tilde x} \right) + \tilde\lambda(x)}] =
 0. \\
 \end{array}
\end{equation}
In (34), $ (a) $ follows from Bayes rule by setting $ p\left(
{\tilde x} \right) = \frac{{p\left( {\left. {\tilde x} \right|x}
\right)p\left( x \right)}}{{p\left( {\left. x \right|\tilde x}
\right)}} $, and, $ (b) $ follows by defining $
\tilde\lambda(x)=\frac{{\lambda \left( x \right)}}{{p\left( x
\right)}}  $. Thus, (34) yields
\begin{equation}
[ {\frac{1}{{\left( {q - 1} \right)}}\left( {\frac{{p\left( {\left.
{\tilde x} \right|x} \right)}}{{p\left( {\tilde x} \right)}}}
\right)^{q - 1}  + \tilde\beta d\left( {x,\tilde x} \right) +
\tilde\lambda(x)} ] =
 0. \\
\end{equation}
 Expanding (35), yields
\begin{equation}
p\left( {\left. {\tilde x} \right|x} \right) = p\left( {\tilde x}
\right)\left[ {\left( {1 - q} \right)\left\{ {\tilde \lambda \left(
x \right) + \tilde\beta d\left( {x,\tilde x} \right)} \right\}}
\right]^{\frac{1}{{q - 1}}}.
\end{equation}

Multiplying the square bracket in (35) by $ p\left( {\left. {\tilde
x} \right|x} \right) $, summing over $ \tilde x $, and, evoking $
\sum\limits_{\tilde x} {p\left( {\left. {\tilde x} \right|X = x}
\right)}  = 1 $ yields
\begin{equation}
\tilde \lambda \left( x \right) = \frac{1}{{\left( {1 - q}
\right)}}\aleph _q \left( x \right) - \tilde\beta \left\langle {d
\left( {x,\tilde x} \right)} \right\rangle _{p\left( {\left. {\tilde
x} \right|X = x} \right)}.
\end{equation}
Here, $ \aleph _q \left( x \right) = \sum\limits_{\tilde x} {p\left(
{\tilde x} \right)\left( {\frac{{p\left( {\left. {\tilde x}
\right|x} \right)}}{{p\left( {\tilde x} \right)}}} \right)^q } $.
Thus (36) yields
\begin{equation}
\begin{array}{l}
 p\left( {\left. {\tilde x} \right|x} \right) = \frac{{p\left( {\tilde x} \right)\left\{ {1 - \left( {q - 1} \right)\beta^* \left( x \right)  d \left( {x,\tilde x} \right)} \right\}^{1/\left( {q - 1} \right)} }}{{(\Im_{RD}(x)) ^{^{1/\left( {1 - q} \right)} } }}, \\

 \Im _{RD} \left( x \right) = \aleph _q \left( x \right) + \left( {q - 1} \right)\tilde\beta \left\langle {d\left( {x,\tilde x} \right)} \right\rangle _{p\left( {\left. {\tilde x} \right|X = x} \right)} =(1-q)\tilde\lambda(x), \\
 \beta^*(x)   = \frac{{ \tilde\beta }}{{\Im _{RD}
\left( x \right) }}
 = \frac{\tilde\beta }{{\aleph _q \left( x \right) + \left( {q - 1} \right)\tilde\beta \left\langle {d\left( {x,\tilde x} \right)} \right\rangle _{p\left( {\left. {\tilde x} \right|X = x} \right)} }}. \\

 \end{array}
\end{equation}
where $ \beta^*(x)  $ is the \textit{\textit{effective nonextensive
trade-off parameter}} for a \textit{single} \textit{source alphabet}
$ x\in X $. The
 \textit{net effective nonextensive trade-off
parameter}, evaluated for all \textit{source alphabets} $ x\in X $
is: $ \beta^* = \sum\limits_x {\beta^*(x)  } $.  Note that the
parameter $ \beta^*(x)  $  explicitly manifests the
\textit{self-referential} nature of the canonical transition
probability $ p(\tilde x|x) $.

The present case differs from the analysis in [21] in the sense that
$ \beta^*(x)  $ is to be evaluated for each \textit{source alphabet}
$ x \in X $. Thus, the so-called \textit{parametric perspective}
employed in [21], by \textit{a-priori} defining a range for $
\beta^*(x) $   $ \in [0,\infty] $ is not possible. Instead the
canonical transition probability $ p(\tilde x|x) $ has to be
evaluated for each $ \beta^*(x) $ for all  $ x \in X $, and, for
each $ \tilde\beta $.  This feature represents a pronounced
qualitative and quantitative distinction when studying the
stationary point solutions of conditional probabilities, as compared
with stationary point solutions of marginal probabilities.

Specifying $ q = 2 - q^* $ in the numerator of (38) and evoking (4),
yields the canonical transition probability
\begin{equation}
p\left( {\left. {\tilde x} \right|x} \right) = \frac{{p\left(
{\tilde x} \right)\exp _{q^ *  } \left[ { - \beta^* \left( x
\right)d\left( {x,\tilde x} \right)} \right]}}{{\sum\limits_{\tilde
x} {p\left( {\tilde x} \right)\exp _{q^ *  } \left[ { - \beta^*
\left( x \right)d\left( {x,\tilde x} \right)} \right]} }} =
\frac{{p\left( {\tilde x} \right)\exp _{q^ *  } \left[ { - \beta^*
\left( x \right)d\left( {x,\tilde x} \right)} \right]}}{{\tilde
Z\left( {x,\beta^* \left( x \right)} \right)}} .
\end{equation}
 The
partition function for a single \textit{source alphabet} $ x \in X $
is
\begin{equation}
\begin{array}{l}
\tilde Z\left( {x,\beta^* \left( x \right)  } \right) = \Im
_{RD}^{\frac{1}{{1 - q}}} \left( x \right) = \sum\limits_{\tilde x}
{p\left( {\tilde x} \right)\exp _{q^*} } \left[ { -  \beta^* \left(
x \right) d \left( {x,\tilde x} \right)} \right].
\end{array}
\end{equation}
Solutions of (39) are only valid for $ \left\{ {1 - \left( {1 - q^*
} \right)  \beta^* \left( x \right)  d\left( {x,\tilde x} \right)}
\right\}
> 0 $, ensuring $ p(\tilde x|x) > 0 $. This is the is the \textit{Tsallis cut-off condition} [1, 21].
The condition $ \left\{ {1 -\left( {1 - q^* } \right) \beta^* \left(
x \right)  d\left( {x,\tilde x} \right)} \right\} \le 0 $ requires
setting $ p(\tilde x|x)=0 $. From (38)-(40), $ \tilde\beta $, $
\beta^* $, and, $ \beta^* \left( x \right) $ relate as
\begin{equation}
\begin{array}{l}
\beta^* \left( x \right) = \frac{\tilde\beta }{{\aleph _q \left( x
\right) + \left( {q - 1} \right)\tilde\beta \left\langle {d\left(
{x,\tilde x} \right)} \right\rangle _{p\left( {\left. {\tilde x}
\right|x} \right)} }} = \frac{\tilde\beta }{{\tilde Z^{\left( {1 -
q} \right)} \left( {x,\beta^* \left( x \right) } \right)}}, \\
\beta^*  = \sum\limits_x {\beta^* \left( x \right) }.
 \end{array}
\end{equation}

\subsection{Case for $ q>1 $}

As discussed in Sections 2 and 3 (Theorem 3) of this paper, use of
the \textit{additive duality} is required to express the generalized
mutual entropy for the range of the nonextensivity parameter $ q
> 1 $ as: $ D^{q^*}_{K-L}[p(X,\tilde X)||p(X)p(\tilde X)] $. This is
done in accordance with the form required for the nonextensive
alternating minimization procedure (31) (Section 4.2) described in $
q^*-space $, and, the convention followed in [3].

\textbf{Lemma 3}: Variational minimization of the Lagrangian
\footnote{Valid for both discrete and continuous cases}

\begin{equation}
\begin{array}{l}
L_{RD}^{q^ *  } \left[ {p\left( {\left. {\tilde x} \right|x}
\right)} \right] = \sum\limits_x {\sum\limits_{\tilde x} {p\left( x
\right)p\left( {\left. {\tilde x} \right|x} \right)} } \frac{{\left(
{\frac{{p\left( {\left. {\tilde x} \right|x} \right)}}{{p\left(
{\tilde x} \right)}}} \right)^{q^ *   - 1}  - 1}}{{q^ *   - 1}} \\
  + \tilde\beta_{q^*} \sum\limits_x {\sum\limits_{\tilde x} {d\left( {x,\tilde x} \right)p\left( x \right)} } p\left( {\left. {\tilde x} \right|x} \right) + \sum\limits_x {\lambda \left( x \right)\sum\limits_{\tilde x} {p\left( {\left. {\tilde x} \right|x} \right)} },  \\

\end{array}
\end{equation}
yields the canonical conditional probability $ p\left( {\left.
{\tilde x} \right|x} \right) = \frac{{p\left( {\tilde x} \right)\exp
_q \left( { - \beta _{2 - q}^
*  \left( x \right)d\left( {x,\tilde x} \right)} \right)}}{{\tilde
Z_{2 - q} \left( {x,\beta _{2 - q}^ *  \left( x \right)} \right)}}
$, where, $\tilde\beta_{q^*}=q^*\beta $,  $\beta^*_{2-q} \left( x
\right) = \frac{\tilde\beta_{q^*} }{{(1-q^*)\tilde \lambda \left( x
\right)}}, \tilde \lambda \left( x \right) = \frac{{\lambda \left( x
\right)}}{{p\left( x \right)}} $, and, $ \sum\limits_x {\sum\limits_{\tilde x} {p\left( {x,\tilde x} \right)}}  = 1 $. \\

\textbf{Proof:}  In accordance with the procedure employed in
Section 5.1, variational minimization of the Lagrangian in (42)
yields
\begin{equation}
\begin{array}{l}
\frac{{\delta L_{RD}^{q^ *  } \left[ {p\left( {\left. {\tilde x}
\right|x} \right)} \right]}}{{\delta p\left( {\left. {\tilde x}
\right|x} \right)}} = p\left( x \right)\left[ {\frac{{1  }}{{q^ * -
1}}\left( {\frac{{p\left( {\left. {\tilde x} \right|x}
\right)}}{{p\left( {\tilde x} \right)}}} \right)^{q^ *   - 1}  +
\tilde\beta_{q^*} d\left( {x,\tilde x} \right) + \tilde \lambda
\left( x
\right)} \right] =0. \\
\end{array}
\end{equation}
Solving (43) for $ p(\tilde x|x) $ and obtaining the normalization
Lagrange multiplier analogous to the approach in Section 5.1, yields
\begin{equation}
\begin{array}{l}
p\left( {\left. {\tilde x} \right|x} \right) = p\left( {\tilde x}
\right)\left[ {\left( {1 - q^ *  } \right)\left\{ {\tilde \lambda
\left( x \right) + \tilde \beta _{q^ *  } d\left( {x,\tilde x}
\right)} \right\}} \right]^{\frac{1}{{q^ *   - 1}}},  \\
and,  \\
 \tilde \lambda \left( x \right) = \frac{{1  }}{{1 - q^ *  }}\underbrace {\sum\limits_{\tilde x} {p\left( {\tilde x} \right)\left( {\frac{{p\left( {\left. {\tilde x} \right|x} \right)}}{{p\left( {\tilde x} \right)}}} \right)^{q^ *  } } }_{\aleph _{q^ *  } \left( x \right)} - \tilde\beta_{q^*}\left\langle { d\left( {x,\tilde x} \right)} \right\rangle _{p\left( {\left. {\tilde x} \right|X = x} \right)}.  \\
\end{array}
\end{equation}

Here, (44) yields
\begin{equation}
\begin{array}{l}
 p\left( {\left. {\tilde x} \right|x} \right) = \frac{{p\left( {\tilde x} \right)\exp _q \left( { - \beta _{2 - q}^ *  \left( x \right)d\left( {x,\tilde x} \right)} \right)}}{{\Im _{RD}^ *  \left( x \right)^{\frac{1}{{q - 1}}} }}, \\
 \Im _{RD}^ *  \left( x \right) = \aleph _{q^ *  } \left( x \right) + (q^*-1)\tilde\beta_{q^*} \left\langle {d
\left( {x,\tilde x} \right)} \right\rangle _{p\left( {\left. {\tilde x} \right|X = x} \right)}, \\
 \beta _{2 - q}^ *  \left( x \right) =
\frac{{\tilde \beta _{q^ *  } }}{{\Im _{RD}^ *  \left( x \right)}} =
\frac{{\tilde \beta _{q^ *  } }}{{\aleph _{q^ *  } \left( x \right)
+ \left( {q^ *   - 1} \right)\tilde \beta _{q^ *  } \left\langle
{d\left( {x,\tilde x} \right)} \right\rangle _{p\left( {\left.
{\tilde x} \right|X = x} \right)} }}. \\
 \end{array}
\end{equation}
Here, (45) yields
  \begin{equation}
  \begin{array}{l}
  p\left( {\left. {\tilde x} \right|x} \right) = \frac{{p\left( {\tilde x} \right)\exp _q \left( { - \beta _{{2 - q}}^ *(x)  d\left( {x,\tilde x} \right)} \right)}}{{\tilde Z_{2 - q} \left( {x,\beta _{{2 - q}}^ *  }(x) \right)}}. \\
 \end{array}
\end{equation}
Solutions of (46) are only valid for $ \left\{ {1 - \left( {1 - q }
\right)\beta ^*_{2-q}(x) d\left( {x,\tilde x} \right)} \right\}
> 0 $, ensuring $ p(\tilde x|x) > 0 $. This is the \textit{Tsallis cut-off condition} [1, 21] for (46), for the range $ 1 <q <2 $.  The condition $ \left\{ {1 -
  \left( {1 - q } \right)\beta ^*_{2-q}(x) d\left( {x,\tilde x} \right)} \right\} \le 0 $ requires setting $ p(\tilde x|x)=0 $.  The partition function is
\begin{equation}
\begin{array}{l}
\tilde Z_{2 - q}\left( {x,\beta _{2 - q}^ * (x) } \right) = \left(
{\Im^* _{RD}  \left( x \right)} \right)^{\frac{1}{{1-q^*}}} =
\sum\limits_{\tilde x} {p\left( {\tilde x} \right)\exp _q \left[ { -
\beta _{2 - q}^ *(x)  d\left( {x,\tilde x} \right)} \right]}.
\end{array}
 \end{equation}

From (45)-(47), $ \tilde\beta_{q^*} $, $ \beta^*_{2-q} $, and, $
\beta^*_{2-q}(x) $ relate as
\begin{equation}
\begin{array}{l}
 \beta _{2 - q}^ *(x)   = \frac{\tilde\beta_{q^*} }{{\aleph _{q^ *  } \left( x \right) + \left( {q^ *   - 1} \right)\tilde\beta_{q^*}\left\langle {d\left( {x,\tilde x} \right)} \right\rangle _{p\left( {\left. {\tilde x} \right|X=x} \right)} }}=\frac{{\tilde \beta _{q^ *  } }}{{\tilde Z_{2 - q}^{\left( {1 - q^ *
} \right)} \left( {x,\beta _{2 - q }^ *(x)  } \right)}}, \\
 \beta _{2 - q}^ *   = \sum\limits_x {\beta _{2 - q}^ *(x)  } . \\
 \end{array}
\end{equation}

\subsection{Nonextensive alternating minimization algorithm revisited}
This sub-Section describes the practical implementation of the
nonextensive alternating minimization algorithm for the case $ 0 < q
< 1 $. This is accomplished using the theory presented in Sections
4.1 and 5.1.  For the sake of brevity, the implementation is
described in point form:

$\bullet $ \textit{A-priori} specifying the nonextensivity parameter
$ q $ and the \textit{effective nonextensive trade-off parameter}
for a single \textit{source alphabet} $ \beta^*(x) $ obtained from
(38) for all \textit{source alphabets}, the expected distortion $ D=
<d(x,\tilde x)>_{p(x,\tilde x)} $ is obtained. Choosing a random
data point in $ \tilde\mathcal{X} $ (the convex set of probability
distributions $ B $, described in Sections 4.1 and 4.2), an initial
guess for $ p(\tilde x) $ is made. Eq. (39) is then employed to
evaluate the transition probability $ p(\tilde x|x) $ (a single data
point in the convex set of probability distributions $ A $,
described in Sections 4.1 and 4.2), that minimizes the generalized
K-Ld $ D^q_{K-L}[\bullet] $ subject to the distortion constraint.

$ \bullet $  Using this value of $ p(\tilde x|x) $, (28) is employed
to calculate a new value of $ p(\tilde x) $ that further minimizes $
D^q_{K-L}[\bullet] $.

$ \bullet $ The above process is repeated thereby monotonically
reducing the right hand side of (31).  Using \textbf{Lemma 2} ((39))
and (21), the algorithm is seen to converge to a unique point on the
RD curve whose slope equals $ -\tilde\beta $.   In principle, for
different values of $ \beta^*(x) $ obtained for all \textit{source
alphabets}, a full RD curve may be obtained.

$ \bullet $ Note that the alternating minimization is performed
independently in the two convex sets of probability distributions $
A $ and $ B $ (see Section 4.2).  Specifically, $ p(\tilde x) $ is
assumed fixed when minimizing with respect to $ p(\tilde x|x) $.  In
the next update step, assuming $ p(\tilde x|x) $ to be fixed, $
p(\tilde x) $ is minimized through (28).

$ \bullet $ In general, the alternating minimization algorithm only
deals with the optimal partitioning of $ \mathcal X $ (induced by $
p(\tilde x|x) $), with respect to a fixed set of representatives ( $
\tilde X $ values).  This implies that the distortion measure $
d(x,\tilde x) $ is pre-defined  and fixed throughout the
implementation $ \forall x \in \mathcal X $ \textit{and} $ \forall
\tilde x \in \tilde \mathcal X $.

The alternating minimization algorithm for nonextensive RD theory is
described in Algorithm 1 for $ 0 < q <1 $.

 \begin{algorithm}
 \caption{Nonextensive Alternating Minimization Scheme for $ 0 < q < 1 $}
 \begin{algorithmic}
 \STATE \textbf{Input} \\

1. Source distribution $ p\left( x \right) \in X $. \\
2. Set of representatives of quantized codebook given by $ p\left(
{\tilde x} \right) \in \tilde X $ values.\\
3. Input \textit{nonextensive trade-off parameter} $ \tilde\beta(=q\beta) $, where, $ \beta $ is the Boltzmann-Gibbs-Shannon \textit{inverse temperature}. \\
4.  Distortion measure $ d(x,\tilde x) $. \\
5. Convergence parameter $ \varepsilon $. \\

\STATE \textbf{Output} \\

Value of $ R_q(D) $ where its slope equals $ - \tilde\beta=-q\beta $.\\

\STATE \textbf{Initialization} \\

Initialize $ R_q^{(0)}  $ and randomly initialize $
p(\tilde x) $ and $ p(\tilde x|x) $ (to initialize $ \beta^{*(0)}(x)
$).

\STATE \textbf{While True} \\

$ \bullet $  $ \beta^ {*(m)}(x)   = \frac{\tilde\beta
}{{\sum\limits_{\tilde x} {p^{(m)}\left( {\tilde x} \right)\left(
{\frac{{p^{(m)}\left( {\left. {\tilde x} \right|x}
\right)}}{{p^{(m)}\left( {\tilde x} \right)}}} \right)^q  + \left(
{q - 1} \right)\tilde\beta \left\langle {d\left( {x,\tilde x} \right)}
\right\rangle _{p^{(m)}\left( {\left. {\tilde x} \right|X = x}
\right)}}
}} $ $ \rightarrow $ \textit{Effective nonextensive trade-off parameter for a single source})(41) \\

 $ \bullet $ $ p^{(m+1)}\left( {\tilde x\left| x \right.} \right)
\leftarrow \frac{{p^{(m)}\left( {\tilde x} \right)\exp _{q^ *  }(
-\beta^{*(m)}(x) d\left( {x,\tilde x}) \right)}}{\tilde
Z^{(m+1)}(x,\beta^{*(m)}) } $
\\
$ \bullet $ $ p^{(m+1)}\left( {\tilde x} \right) \leftarrow
\sum\limits_x {p\left( x \right)} p^{(m+1)}\left( {\tilde x\left| x
\right.} \right) $
\\
$ R_q^{\left( {m + 1} \right)} \left( D \right) = D^q_{K-L} \left[
{p\left( x \right)p^{\left( {m + 1} \right)} \left( {\left. {\tilde
x} \right|x} \right)||p\left( x \right)p^{\left( {m + 1} \right)}
\left( {\tilde x} \right)} \right] $.
\\
If $ \left( {R_q^{\left( m \right)} \left( D \right) - R_q^{\left(
{m + 1} \right)} \left( D \right)} \right) \le \varepsilon $
\\

Break \\

$\bullet $ Test Tsallis cut-off condition.\\
$ \bullet $ $ \tilde\beta \leftarrow \tilde\beta+\delta\tilde\beta $

\end{algorithmic}
\end{algorithm}

\section{Numerical simulations and physical interpretations}

The qualitative distinctions between nonextensive statistics and
extensive statistics is demonstrated with the aid of the respective
RD models.  To this end, a sample of 500 two-dimensional data points
is drawn from three spherical Gaussian distributions with means at $
(2, 3.5),(0, 0),(0,2) $ (the \textit{quantized codebook}). The
priors and standard deviations are $ 0.3,0.4,0.3 $, and, $
0.2,0.5,1.0 $, respectively. The distortion measure $ d(x,\tilde x)
$ is taken to be the Euclidean square distance. The case $ 0 <q <1 $
(Section 5.1) is chosen for the numerical study.  The axes of the
nonextensive RD curves are scaled with respect to those of the
extensive Boltzmann-Gibbs-Shannon RD curve.  The nonextensive RD
numerical simulations are performed for three values of the
nonextensivity parameter $ (i) $ $ q=0.85 $, $ (ii) $ $ q=0.70 $,
and, $ (iii) $ $ q=0.5 $, respectively.  \textit{The nonextensive RD
model demonstrates extreme sensitivity to the problem size (model
complexity) and the nature of the test data}.

Fig.2 depicts the extensive and nonextensive RD curves. Each curve
has been generated for values of the Boltzmann-Gibbs-Shannon
\textit{inverse temperature} $ \beta \in \left[ {.1,2.2} \right] $.
Note that in this paper, the \textit{nonextensive trade-off
parameter} is: $ \tilde\beta=q\beta$.
 \textit{It is observed that the nonextensive RD theory exhibits a lower threshold
for the minimum achievable compression-information in the
\textit{distortion-compression plane}, as compared to the extensive
case, for all values of $ q $ in the range $ \ 0 < q < 1 \ $ }.
The nonextensive RD curves are \textit{upper bounded} by the Boltzmann-Gibbs-Shannon RD curve.

Specifically, as described in Section 5.3 of this paper, first the
transition probability $ p(\tilde x|x) $ is obtained by solving
(39), followed by a correction of the marginal probability $
p(\tilde x) $ using (24). This process is iteratively applied for
each individual value of $ \tilde\beta $ till convergence is
reached.  The value of $ \tilde\beta $ is then marginally increased,
resulting in the nonextensive RD curve.  This is the crux of the
nonextensive alternating minimization procedure. The nonextensive RD
curves in Fig. 2 are truncated, by terminating the nonextensive
alternating minimization algorithm when the \textit{Tsallis cut-off
condition} is breached.

Note that for the nonextensive cases, the slope of the tangent drawn
at any point on the nonextensive RD curve is the negative of the
nonextensive trade-off parameter $ -\tilde\beta=-q\beta $. Data
clustering may be construed as being a form of lossy data
compression [40].  At the commencement, the nonextensive alternating
minimization algorithm solves for the compression phase with $
\tilde\beta  \to 0 $. The compression phase is characterized by all
data points "coalescing" around a single data point $ \tilde x \in
\tilde X $, in order to achieve the most compact representation ($
R_q(D) \to 0 $).

As $ \tilde\beta $ increases, the data points undergo
\textit{soft clustering} around the cluster centers. By definition, in \textit{soft clustering},
a data point $ x \in X $ is assigned to a given cluster whose
centers are $ \tilde x \in \tilde X $ through a normalized
transition probability $ p(\tilde x|x) $. The \textit{hard
clustering} regime signifies regions where $ \tilde\beta \to \infty $.  By definition, in \textit{hard
clustering} the assignment of data points to clusters is
deterministic.

An observation of particular significance is revealed in Fig.2.
Specifically, even for less relaxed distortion constraints $
\left\langle {d\left( {x,\tilde x} \right)} \right\rangle _{p\left(
{x,\tilde x} \right)} $, any nonextensive case for $ \ 0 < q < 1 \ $
possesses a lower minimum \textit{compression information }than the
corresponding extensive case.  The threshold for the minimum
achievable \textit{compression information} $ I_q(X;\tilde X) $
decreases as $ q \to 0 $. \textit{Note that all nonextensive RD
curves inhabit the non-achievable region for the extensive case}. By
definition, the \textit{non-achievable region} is the region below a
given RD curve, and signifies the domain in the
\textit{distortion-compression} plane where compression does not
occur.

\textit{Further, nonextensive RD models possessing a lower
nonextensivity parameter $ q $ inhabit the non-achievable regions of
nonextensive RD models possessing a higher value of $ q $.}  These
features imply the superiority of nonextensive models to perform
data compression \textit{vis-\'{a}-vis} any comparable model derived
from Boltzmann-Gibbs-Shannon statistics.

\section{Summary and conclusions}

Variational principles for a generalized RD theory, and, a \textit{dual}
generalized RD theory employing the \textit{additive duality} of nonextensive
statistics, have been presented.  This has been accomplished using a
methodology to "rescue" the linear constraints originally employed
by Tsallis [1], formulated in [21]. Select information theoretic
properties of \textit{dual} Tsallis uncertainties have been investigated
into. Numerical simulations have proven that the nonextensive RD
models demonstrate a lower threshold for the \textit{compression
information} \textit{vis-\'{a}-vis} equivalent models derived from
Boltzmann-Gibbs-Shannon statistics.  This feature acquires
significance in data compression applications.

The nonextensive RD models and the nonextensive alternating
minimization numerical scheme studied in this paper represent
idealized scenarios, involving well behaved sources and distortion
measures. Based on the results reported herein, an ongoing study has
treated a more realistic generalized RD scenario by extending the
works of Rose [41] and Banerjee \textit{et. al.} [42], and has
accomplished a three-fold objective.

First, a generalized Bregman RD (GBRD) model has been formulated
using the nonextensive alternating minimization algorithm as its
basis. Let $ \tilde \mathcal {X}_s $ be a subset of $ \tilde
\mathcal X $, where $ p(\tilde x) \neq 0 $ (\textit{the support}).
From a computational viewpoint, the GBRD model represents a
\textit{non-convex optimization problem}, where the cardinality of $
\tilde \mathcal {X}_s $ ( $ |\tilde \mathcal {X}_s | $) varies with
increase in the \textit{nonextensive trade-off parameter}.     Next
a \textit{Tsallis-Bregman lower bound} for the RD function is
derived. The \textit{Tsallis-Bregman lower bound} provides a
principled theoretical rationale for the lower threshold for the
\textit{compression information} demonstrated by generalized
statistics RD models, \textit{vis-\'{a}-vis} equivalent extensive RD
models.

Finally, the problem of rate distortion in lossy data compression is
shown to be equivalent to the problem of mixture model estimation in
unsupervised learning [43].  This is demonstrated for
\textit{q-deformed} exponential families of distributions [44, 45].
The primary rationale  for this exercise is to solve the
 generalized RD problem employing an Expectation-Maximization\textit{-like}
 algorithm [39], using the results of Lemma 1 as a basis.  Results of these studies will be presented elsewhere.

\textbf{Acknowledgements}

RCV gratefully acknowledges support from \textit{RAND-MSR} contract
\textit{CSM-DI $ \ \& $ S-QIT-101107-2005}.  Comments from the
anonymous referees are gratefully acknowledged.

\newpage

\section*{FIGURE CAPTIONS}

\textbf{Fig. 1:}  Schematic diagram for rate distortion curve.
\\
\textbf{Fig. 2:}  Rate distortion curves.
Boltzmann-Gibbs-Shannon model (solid line), generalized statistics RD model for
$ q=0.85 $ (dash-dots), $ q=0.70 $ (dashes), and, $ q=0.5 $ (dots).
\\
\newpage

\begin{figure}[thpb]
\centering
\begin{center}
\includegraphics[scale=1.00]{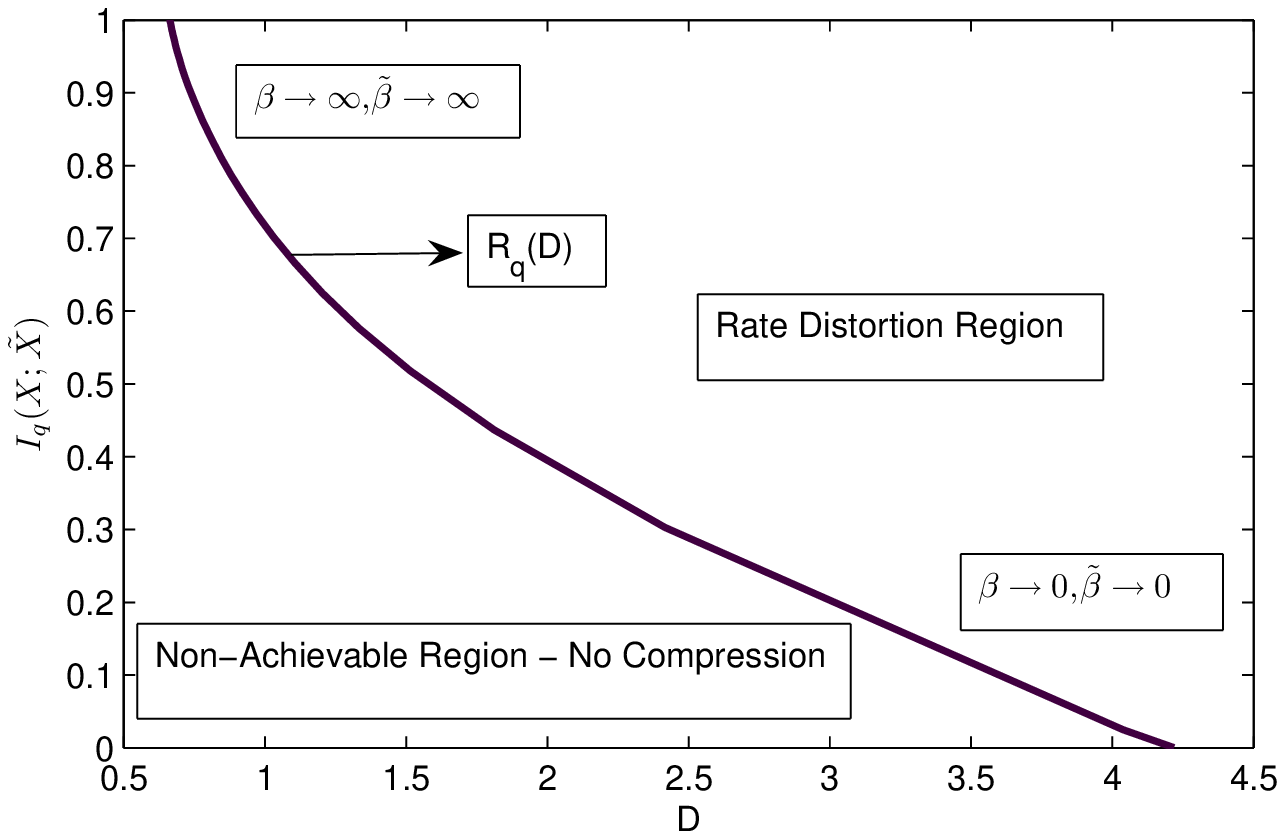}
\end{center}
\end{figure}

\newpage

\begin{figure}[thpb]
\centering
\begin{center}
\includegraphics[scale=1.00]{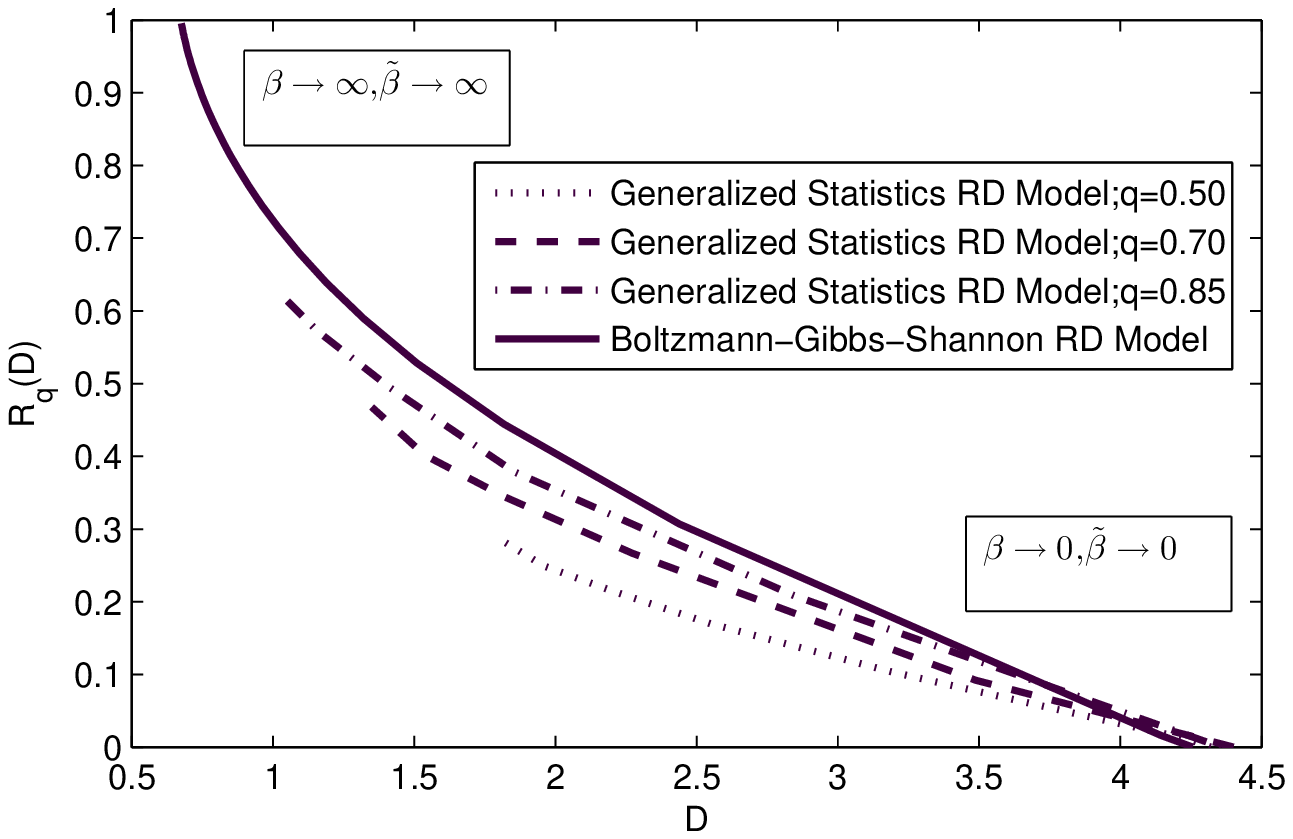}
\end{center}
\end{figure}

\end{document}